\def\year{2021}\relax
\documentclass[letterpaper]{article} 
\usepackage{aaai21}  
\usepackage{times}  
\usepackage{helvet} 
\usepackage{courier}  
\usepackage[hyphens]{url}  
\usepackage{graphicx} 
\usepackage{natbib}  
\usepackage{caption} 

\usepackage{subfiles}
\usepackage{subfigure}
\usepackage{booktabs} 
\usepackage{epsfig,endnotes}
\usepackage{multirow}
\usepackage{xspace} 
\usepackage{mathrsfs}
\usepackage{bm}
\usepackage{amssymb}
\usepackage{amsmath}
\usepackage{amsthm}
\usepackage{epstopdf}
\usepackage{enumitem}
\usepackage{fmtcount}
\usepackage{sectsty}
\newtheorem{theorem}{Theorem}
\usepackage{thm-restate,thmtools}

\newcommand{\myparatight}[1]{\smallskip\noindent{\bf {#1}:}~}
\newcommand{\argmax}{\operatornamewithlimits{argmax}}

\allowdisplaybreaks
\usepackage{algorithm}
\usepackage{algorithmic}
\usepackage{makecell}
\newcommand{\RomanNumeralCaps}[1]
    {\MakeUppercase{\romannumeral #1}}

\title{Provably Secure Federated Learning against Malicious Clients}
\author{
       Xiaoyu Cao,
       Jinyuan Jia, 
       Neil Zhenqiang Gong \\
}
\affiliations {
    Duke University \\
    \{xiaoyu.cao, jinyuan.jia, neil.gong\}@duke.edu
}

\begin{document}

\maketitle


\begin{abstract}
Federated learning enables clients to collaboratively learn a shared global model without sharing their local training data with a cloud server. However, malicious clients can corrupt the global model to predict incorrect labels for testing examples. Existing defenses against malicious clients leverage Byzantine-robust federated learning methods. However, these methods cannot provably guarantee that the predicted label for a testing example is  not affected by malicious clients.  We bridge this gap via \emph{ensemble federated learning}.  In particular, given any base federated learning algorithm, we use the algorithm to learn multiple global models, each of which is learnt using a randomly selected subset of clients. When predicting the label of a testing example, we take majority vote among the  global models. We show that our ensemble federated learning with any base federated learning algorithm is provably secure against malicious clients. Specifically, the label predicted by our ensemble global model for a testing example is provably not affected by a bounded number of malicious clients. Moreover, we show that our derived bound is tight. We evaluate our method on MNIST and Human Activity Recognition datasets. For instance, our method can achieve a certified accuracy of 88\% on MNIST when 20 out of 1,000 clients are malicious.
\end{abstract}


\section{Introduction}

Federated learning \cite{konevcny2016federated,mcmahan2016communication} is an emerging machine learning paradigm, which enables many clients (e.g., smartphones, IoT devices, and organizations) to collaboratively learn a model without sharing their local training data with a cloud server. Due to its promise for protecting privacy of the clients' local training data and the emerging privacy regulations such as General Data Protection Regulation (GDPR), federated learning has been deployed by industry. For instance, Google has deployed  federated learning for next-word prediction on Android Gboard.  Existing federated learning methods mainly follow a \emph{single-global-model} paradigm.  Specifically, a cloud server maintains a \emph{global model} and each client maintains a \emph{local model}.  The global model is trained via multiple iterations of communications between the clients and server. In each iteration, three steps are performed: 1) the server sends the current global model to the clients; 2) the clients update their local models based on the global model and their local training data, and send the model updates to the server; and 3) the server aggregates the model updates and uses them to update the global model. The learnt global model is then used to predict labels of testing examples.

However, such single-global-model paradigm is vulnerable to security attacks. In particular, an attacker can inject fake clients to federated learning or compromise existing clients, where we call the fake/compromised clients \emph{malicious clients}. Such malicious clients can corrupt the global model via carefully tampering their local training data or model updates sent to the server. As a result, the corrupted global model has a low accuracy for the normal testing examples \cite{fang2019local,Xie19} or certain attacker-chosen testing examples~\cite{Bagdasaryan18,Bhagoji19,xie2019dba}. For instance, when learning an image classifier, the malicious clients can re-label the cars with certain strips as birds in their local training data and scale up their model updates sent to the server, such that the learnt global model incorrectly predicts a car with the strips as bird \cite{Bagdasaryan18}. 

Various Byzantine-robust federated learning methods have been proposed to defend against malicious clients \cite{Blanchard17,ChenPOMACS17,Mhamdi18,Yin18,yin2019defending,chen2018draco,alistarh2018byzantine}. The main idea of these methods  is to mitigate the impact of statistical outliers among the clients' model updates. They can bound the difference between the global model parameters learnt without malicious clients and the global model parameters learnt when some clients become malicious. However, these methods cannot provably guarantee that the label predicted by the global model for a testing example is not affected by malicious clients. Indeed, studies showed that malicious clients can still substantially degrade the testing accuracy of a global model learnt by a Byzantine-robust method via carefully tampering their model updates sent to the server~\cite{Bhagoji19,fang2019local,Xie19}. 

In this work, we propose \emph{ensemble federated learning}, the first federated learning method that is provably secure against malicious clients. Specifically, given $n$ clients, we define a \emph{subsample} as a set of $k$ clients sampled from the $n$ clients uniformly at random without replacement. For each subsample, we can learn a global model using a base federated learning algorithm with the $k$ clients in the subsample. Since there are ${n \choose k}$ subsamples with $k$ clients, ${n \choose k}$ global models can be trained in total. Suppose we are given a testing example $\mathbf{x}$. We define $p_i$ as the fraction of the ${n \choose k}$ global models that predict label $i$ for $\mathbf{x}$, where $i=1,2,\cdots,L$. We call $p_i$ \emph{label probability}. Our \emph{ensemble global model} predicts the label with the largest label probability for $\mathbf{x}$. In other words, our ensemble global model takes a majority vote among the global models to predict label for $\mathbf{x}$. Since each global model is learnt using a subsample with $k$ clients, a majority of the global models are learnt using normal clients when most clients are normal. Therefore, the majority vote among the global models is secure against a bounded number of malicious clients.

\myparatight{Theory} Our first major theoretical result is that our ensemble global model provably predicts the same label for a testing example $\mathbf{x}$ when the number of malicious clients is no larger than a threshold, which we call \emph{certified security level}.  Our second major theoretical result is that we prove our derived certified security level is tight, i.e., when no assumptions are made on the base federated learning algorithm, it is impossible to derive a certified security level that is larger than ours. Note that the certified security level may be different for different testing examples. 

\myparatight{Algorithm} Computing our certified security level for $\mathbf{x}$ requires its largest and second largest label probabilities. When ${n \choose k}$ is small (e.g., the $n$ clients are dozens of organizations \cite{kairouz2019advances} and $k$ is small), we can compute the largest and second largest label probabilities exactly via training ${n \choose k}$ global models. 
However, it is challenging to compute them exactly when ${n \choose k}$ is large. To address the computational challenge, we develop a Monte Carlo algorithm to estimate them with probabilistic guarantees via training $N$ instead of ${n \choose k}$ global models.  

\myparatight{Evaluation} We empirically evaluate our method on MNIST \cite{lecun2010mnist} and Human Activity Recognition datasets \cite{anguita2013public}. We distribute the training examples in MNIST to clients to simulate  federated learning scenarios, while the Human Activity Recognition dataset represents a real-world federated learning scenario, where each user is a client. We use the popular FedAvg developed by Google~\cite{mcmahan2016communication} as the base federated learning algorithm. Moreover, we use \emph{certified accuracy} as our evaluation metric, which is a lower bound of the testing accuracy that a method can provably achieve no matter how the malicious clients tamper their local training data and model updates. For instance, our ensemble FedAvg with $N=500$ and $k=10$ can achieve a certified accuracy of 88\% on MNIST when evenly distributing the training examples among 1,000 clients and 20 of them are malicious.  

In summary, our key contributions are as follows:
\begin{itemize}
\item {\bf Theory:} We propose ensemble federated learning, the first provably secure federated learning method against malicious clients. We derive a certified security level for our ensemble federated learning. Moreover, we prove that our derived certified security level is tight.
\item {\bf Algorithm:} We propose a Monte Carlo algorithm to compute our certified security level in practice. 
\item {\bf Evaluation:} We evaluate our methods on MNIST and Human Activity Recognition datasets.
\end{itemize}

All our proofs are shown in Supplemental Material.


\section{Background on Federated Learning}

\begin{algorithm}[t]
	\caption{Single-global-model federated learning}\label{alg:base_alg}
	\begin{algorithmic}[1]
		\STATE {\bfseries Input:} $\mathbf{C}$, $globalIter$, $localIter$, $\eta$, Agg.
    \STATE {\bfseries Output:} Global model $\mathbf{w}$.\\
		\STATE $\mathbf{w} \leftarrow $ random initialization.
		\FOR{$Iter\_global=1,2,\cdots,globalIter$}
		    \STATE /* Step \RomanNumeralCaps{1} */ 
		    \STATE The server sends $\mathbf{w}$ to the clients.
	        \STATE /* Step \RomanNumeralCaps{2} */ 
		    \FOR{$i\in \mathbf{C}$}
		        \STATE $\mathbf{w}_i \leftarrow \mathbf{w}$.
		        \FOR{$Iter\_local=1,2,\cdots,localIter$}
		            \STATE Sample a $Batch$ from local training data $\mathcal{D}_i$.
		            \STATE $\mathbf{w}_i \leftarrow \mathbf{w}_i - \eta \nabla Loss(Batch;\mathbf{w}_i)$.
		        \ENDFOR
	            \STATE Send $\mathbf{g}_i=\mathbf{w}_i-\mathbf{w}$ to the server.
		    \ENDFOR
		    \STATE /* Step \RomanNumeralCaps{3} */ 
            \STATE $\mathbf{g} \leftarrow \text{Agg}(\mathbf{g}_1, \mathbf{g}_2, \cdots, \mathbf{g}_{|\mathbf{C}|})$.
            \STATE $\mathbf{w} \leftarrow \mathbf{w} - \eta\cdot \mathbf{g}$.
		\ENDFOR\\
		\RETURN $\mathbf{w}$.
	\end{algorithmic} 
\end{algorithm}

Assuming we have $n$ clients $\mathbf{C}=\{1,2,\cdots,n\}$ and a cloud server in a federated learning setting. The $i$th client holds some local training dataset $\mathcal{D}_i$, where $i=1,2,\cdots,n$. Existing federated learning methods \cite{konevcny2016federated,mcmahan2016communication,wang2020federated,li2019convergence} mainly focus on learning a single global model for the $n$ clients. 
Specifically, the server maintains a global model and each client maintains a local model. Then, federated learning iteratively performs the following three steps, which are shown in Algorithm \ref{alg:base_alg}. In Step I, the server sends the current global model to the clients.\footnote{The server may select a subset of clients, but we assume the server sends the global model to all clients for convenience.} In Step II, each client trains a local model via fine-tuning the global model to its local training dataset. In particular, each client performs \emph{localIter} iterations of stochastic gradient descent with a learning rate $\eta$ to train its local model. Then, each client sends its model update (i.e., the difference between the local model and the global model) to the server. In Step III, the server aggregates the clients' model updates according to some aggregation rule \emph{Agg} and uses the aggregated model update to update the global model. The three steps are repeated for \emph{globalIter} iterations. Existing federated learning algorithms essentially use different aggregation rules in Step III. For instance, Google developed FedAvg \cite{mcmahan2016communication}, which computes the average of the clients' model updates weighted by the sizes of their local training datasets as the aggregated model update  to update the global model. 

We call such a federated learning algorithm that learns a single global model  \emph{base federated learning algorithm} and denote it as $\mathcal{A}$. Note that given any subset of the $n$ clients $\mathbf{C}$, a base federated learning algorithm can learn a global model for them. Specifically, the server learns a global model via iteratively performing the three steps between the server and the given subset of clients. 


\section{Our Ensemble Federated Learning}

Unlike single-global-model federated learning, our ensemble federated learning trains multiple global models, each of which is trained using the base algorithm $\mathcal{A}$ and a subsample with $k$ clients sampled from the $n$ clients uniformly at random without replacement. Among the $n$ clients $\mathbf{C}$, we have ${n \choose k}$ subsamples with $k$ clients. Therefore,   ${n \choose k}$ global models can be trained in total if we train a global model using each subsample. For a given testing input $\mathbf{x}$, these global models may predict different labels for it. We define $p_i$ as the fraction of the ${n \choose k}$ global models that predict label $i$ for $\mathbf{x}$, where $i=1,2,\cdots, L$. We call $p_i$ \emph{label probability}. Note that $p_i$ is an integer multiplication of  $\frac{1}{{n \choose k}}$, which we will leverage to derive a tight security guarantee of ensemble federated learning. Moreover, $p_i$ can also be viewed as the probability that a global model trained on a random subsample with $k$ clients predicts label $i$ for $\mathbf{x}$. 
 Our \emph{ensemble global model} predicts the label with the largest label probability for $\mathbf{x}$, i.e., we define:
\begin{equation}
    h(\mathbf{C},\mathbf{x}) = \argmax_i p_i,
    \label{eq:define_k}
\end{equation}
where $h$ is our {ensemble global model} and $h(\mathbf{C}, \mathbf{x})$ is the label that our {ensemble global model} predicts for $\mathbf{x}$ when the ensemble global model is trained on clients $\mathbf{C}$. 

\myparatight{Defining provable security guarantees against malicious clients} 
Suppose some of the $n$ clients $\mathbf{C}$ become malicious. These malicious clients can arbitrarily tamper their local training data and model updates sent to the server in each iteration of federated learning. We denote by $\mathbf{C}'$ the set of $n$ clients with malicious ones. Moreover, we denote by $M(\mathbf{C}')$ the number of malicious clients in $\mathbf{C}'$, e.g., $M(\mathbf{C}')= m$ means that $m$ clients are malicious. Note that we don't know which clients are malicious. For a testing example $\mathbf{x}$, our goal is to show that our ensemble global model $h$ provably predicts the same label for $\mathbf{x}$ when the number of malicious clients is bounded. Formally, we aim to show the following:
\begin{align}
\label{securitylevel}
h(\mathbf{C}',\mathbf{x})=h(\mathbf{C},\mathbf{x}), \forall \mathbf{C}', M(\mathbf{C}') \leq m^*,
\end{align}
where $h(\mathbf{C}',\mathbf{x})$ is the label that the ensemble global model trained on the clients $\mathbf{C}'$ predicts for $\mathbf{x}$. We call $m^*$ \emph{certified security level}. When a global model satisfies Equation (\ref{securitylevel}) for a testing example $\mathbf{x}$, we say the global model achieves a provable security guarantee for $\mathbf{x}$ with a certified security level $m^*$. Note that the certified security level may be different for different testing examples. Next, we derive the certified security level of our ensemble global model. 

\myparatight{Deriving certified security level using exact label probabilities} 
Suppose we are given a testing example $\mathbf{x}$. Assuming that, when there are no malicious clients, our ensemble global model predicts label $y$ for $\mathbf{x}$, $p_y$ is the largest label probability, and $p_z$ is the second largest label probability. 
Moreover, we denote by $p_y'$ and $p_z'$ respectively the label probabilities for $y$ and $z$ in the ensemble global model when there are malicious clients. Suppose $m$ clients become malicious. Then, $1 - \frac{{n-m \choose k}}{{n \choose k}}$ fraction of subsamples with $k$ clients include at least one malicious client. In the worst-case scenario, for each global model learnt using a subsample including at least one malicious client, its predicted label for $\mathbf{x}$ changes from $y$ to $z$. Therefore, in the worst-case scenario, the $m$ malicious clients decrease the largest label probability $p_y$ by $1 - \frac{{n-m \choose k}}{{n \choose k}}$ and increase the second largest label probability $p_z$ by $1 - \frac{{n-m \choose k}}{{n \choose k}}$, i.e., we have $p_y'=p_y-(1 - \frac{{n-m \choose k}}{{n \choose k}})$ and $p_z'=p_z + (1 - \frac{{n-m \choose k}}{{n \choose k}})$. Our ensemble global model still predicts label $y$ for $\mathbf{x}$, i.e., $h(\mathbf{C}',\mathbf{x})=h(\mathbf{C},\mathbf{x})=y$, once $m$ satisfies the following inequality:
\begin{align}
\label{inequalitycondition}
p_y' > p_z' 
\Longleftrightarrow p_y-p_z > 2 - 2\frac{{n-m \choose k}}{{n \choose k}}.
\end{align}
In other words, the largest integer $m$ that satisfies the inequality (\ref{inequalitycondition}) is our certified security level $m^*$ for the testing example $\mathbf{x}$. The inequality (\ref{inequalitycondition}) shows that our certified security level is related to the gap $p_y-p_z$ between the largest and second largest label probabilities in the ensemble global model trained on the clients $\mathbf{C}$ without malicious ones. For instance, when a testing example has a larger  gap $p_y-p_z$, the inequality (\ref{inequalitycondition}) may be satisfied by a larger $m$, which means that our ensemble global model may have a larger certified security level for the testing example. 

\myparatight{Deriving certified security level using approximate label probabilities} When ${n\choose k}$ is small (e.g., several hundred), we can compute the exact label probabilities $p_y$ and $p_z$ via training ${n\choose k}$ global models, and compute the certified security level via inequality (\ref{inequalitycondition}). However, when ${n\choose k}$ is large,  it is computationally challenging to compute the exact label probabilities via training ${n\choose k}$ global models. For instance, when $n=100$ and $k=10$, there are already $1.73\times 10^{13}$ global models, training all of which is computationally intractable in practice. Therefore, we also derive certified security level using a lower bound $\underline{p_y}$ of $p_y$ (i.e., $\underline{p_y}\leq p_y$) and an upper bound $\overline{p}_z$  of $p_z$ (i.e., $\overline{p}_z\geq p_z$). We use a lower bound $\underline{p_y}$ of $p_y$ and an upper bound $\overline{p}_z$  of $p_z$ because our certified security level is related to the gap $p_y-p_z$ and we aim to estimate a lower bound of the gap. The lower bound $\underline{p_y}$ and upper bound $\overline{p}_z$ may be estimated by different methods. For instance, in the next section, we propose a Monte Carlo algorithm to estimate a lower bound $\underline{p_y}$ and an upper bound $\overline{p}_z$  via only training $N$ of the ${n\choose k}$ global models. 

\begin{figure}[!t]
    \center
   {\includegraphics[width=0.4\textwidth]{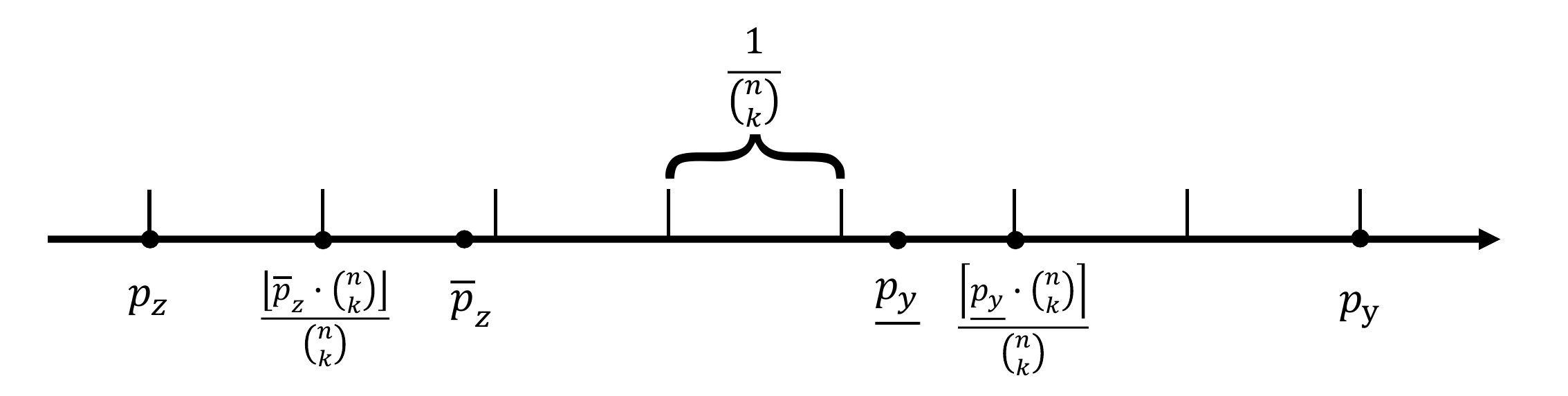}}  
   \vspace{-3mm}
    \caption{An example to illustrate the relationships between $p_y, \underline{p_y}$, and  $\frac{\left\lceil\underline{p_y} \cdot {{n \choose k}}\right\rceil}{{n \choose k}}$ as well as  $p_z, \overline{p}_z$, and $\frac{\left\lfloor\overline{p}_z \cdot {n \choose k}\right\rfloor}{{n\choose k}}$.}
    \label{fig:remark}
\vspace{-3mm}
\end{figure}

Next, we derive our certified security level based on the probability bounds $\underline{p_y}$ and $\overline{p}_z$. 
One way  is to replace $p_y$ and $p_z$ in inequality (\ref{inequalitycondition}) as $\underline{p_y}$ and $\overline{p}_z$, respectively. Formally, we have the following inequality: 
\begin{align}
\label{inequalitycondition1}
 \underline{p_y}-\overline{p}_z > 2 - 2\frac{{n-m \choose k}}{{n \choose k}}.
\end{align}
If an $m$ satisfies inequality (\ref{inequalitycondition1}), then the $m$ also satisfies inequality (\ref{inequalitycondition}), because  $\underline{p_y}-\overline{p}_z \leq p_y-p_z$. Therefore, we can find the largest integer $m$ that satisfies the inequality (\ref{inequalitycondition1}) as the certified security level $m^*$. However, we found that the certified security level $m^*$ derived based on inequality (\ref{inequalitycondition1}) is not tight, i.e., our ensemble global model may still predict label $y$ for $\mathbf{x}$ even if the number of malicious clients is larger than $m^*$ derived based on inequality (\ref{inequalitycondition1}). The key reason is that the label probabilities are integer multiplications of $\frac{1}{{n \choose k}}$. Therefore, we normalize $\underline{p_y}$ and $\overline{p}_z$ as integer multiplications of $\frac{1}{{n \choose k}}$ to derive a tight certified security level. Specifically, we derive the certified security level as the largest integer $m$ that satisfies the following inequality (formally described in Theorem~\ref{certified_radius}):
\begin{align}
\label{eq:certified_condition1}
\frac{\left\lceil\underline{p_y} \cdot {n \choose k}\right\rceil}{{n \choose k}} - \frac{\left\lfloor\overline{p}_z \cdot {n \choose k}\right\rfloor}{{n\choose k}}> 2 - 2\cdot \frac{{n-m \choose k}}{{n \choose k}}.
\end{align}
Figure~\ref{fig:remark} illustrates the relationships between $p_y, \underline{p_y}$, and  $\frac{\left\lceil\underline{p_y} \cdot {{n \choose k}}\right\rceil}{{n \choose k}}$ as well as  $p_z, \overline{p}_z$, and $\frac{\left\lfloor\overline{p}_z \cdot {n \choose k}\right\rfloor}{{n\choose k}}$. When an $m$ satisfies  inequality (\ref{inequalitycondition1}), the $m$ also satisfies  inequality (\ref{eq:certified_condition1}), because $\underline{p_y}-\overline{p}_z \leq \frac{\left\lceil\underline{p_y} \cdot {{n \choose k}}\right\rceil}{{n \choose k}} - \frac{\left\lfloor\overline{p}_z \cdot {n \choose k}\right\rfloor}{{n\choose k}}$. Therefore, the certified security level derived based on inequality (\ref{inequalitycondition1}) is smaller than or equals the certified security level derived based on inequality (\ref{eq:certified_condition1}). Note that when $\underline{p_y}=p_y$ and $\overline{p}_z=p_z$, both (\ref{inequalitycondition1}) and (\ref{eq:certified_condition1}) reduce to (\ref{inequalitycondition}) as the label probabilities are integer multiplications of $\frac{1}{{n \choose k}}$.  The following  theorem formally summarizes our certified security level.  

\begin{theorem}
\label{certified_radius}
Given $n$ clients $\mathbf{C}$, an arbitrary base federated learning algorithm $\mathcal{A}$,  a subsample size $k$, and a testing example $\mathbf{x}$, we define an ensemble global model $h$ as Equation (\ref{eq:define_k}). $y$ and $z$ are the labels that have the largest and second largest label probabilities for $\mathbf{x}$ in the ensemble global model. $\underline{p_y}$ is a lower bound of $p_y$ and $\overline{p}_z$ is an upper bound of $p_z$. Formally, $\underline{p_y}$ and $\overline{p}_z$ satisfy the following conditions:
\begin{equation}
\label{eq:prob_condition}
    \max_{i\neq y} p_i = p_z \le \overline{p}_z \le \underline{p_y} \le p_y.
\end{equation} 
Then, $h$ provably predicts $y$ for  $\mathbf{x}$ when at most $m^*$ clients in $\mathbf{C}$ become malicious, i.e., we have:
{\small
\begin{align}
      \;h(\mathbf{C'},  \mathbf{x}) =h(\mathbf{C}, \mathbf{x}) =y, \forall \mathbf{C}', M(\mathbf{C}') \leq m^*,
\end{align}
}
where $m^*$ is the largest integer $m$ ($0 \le  m \le n-k$) that satisfies inequality (\ref{eq:certified_condition1}). 
\end{theorem}

Our Theorem~\ref{certified_radius} is applicable to any base federated learning algorithm, any  lower bound $\underline{p_y}$ of $p_y$ and any upper bound $\overline{p}_z$ of $p_z$ that satisfy (\ref{eq:prob_condition}). When the lower bound $\underline{p_y}$ and upper bound $\overline{p}_z$ are estimated more accurately, i.e., $\underline{p_y}$ and $\overline{p}_z$ are respectively closer to $p_y$ and $p_z$, our certified security level may be larger. 
The following theorem shows that our derived certified security level is tight, i.e., when no assumptions on the base federated learning algorithm are made, it is impossible to derive a certified security level that is larger than ours for the given probability bounds $\underline{p_y}$ and $\overline{p}_z$. 
 
\begin{theorem}
\label{tightness_theorem}
Suppose $\underline{p_y} + \overline{p}_z \le 1$. For any $\mathbf{C'}$ satisfying $M(\mathbf{C'})>m^*$, i.e., at least $m^*+1$ clients are malicious, there exists a base federated learning algorithm $\mathcal{A}^*$ that satisfies (\ref{eq:prob_condition}) but $h(\mathbf{C'}, \mathbf{x}) \neq y$ or there exist ties.
\end{theorem}


\section{Computing the Certified Security Level}
Suppose we are given $n$ clients $\mathbf{C}$, a base federated learning algorithm $\mathcal{A}$, a subsample size $k$, and a testing dataset $\mathcal{D}$ with $d$ testing examples. For each testing example $\mathbf{x}_t$ in $\mathcal{D}$, we aim to compute its label $\hat{y}_t$ predicted by our ensemble global model $h$ and the corresponding certified security level $\hat{m}_t^*$. To compute the certified security level based on our Theorem~\ref{certified_radius}, we need a lower bound $\underline{p_{\hat{y}_t}}$ of the largest label probability ${p_{\hat{y}_t}}$ and an upper bound $\overline{p}_{\hat{z}_t}$ of the second largest label probability ${p}_{\hat{z}_t}$. When ${n \choose k}$ is small, we can compute the exact label probabilities via training ${n \choose k}$ global models.  When ${n \choose k}$ is large, we propose a Monte Carlo algorithm to  estimate the predicted label and the two probability bounds for all testing examples in $\mathcal{D}$ simultaneously with a confidence level $1-\alpha$ via training $N$ of the ${n \choose k}$ global models. 

\myparatight{Computing predicted label and probability bounds for one testing example} We first discuss how to compute the predicted label $\hat{y}_t$ and probability bounds $\underline{p_{\hat{y}_t}}$ and $\overline{p}_{\hat{z}_t}$ for one testing example $\mathbf{x}_t$. We sample $N$ subsamples with $k$ clients from the $n$ clients uniformly at random without replacement and use them to train $N$ global models $g_1, g_2, \cdots, g_{N}$. We use the $N$ global models to predict labels for $\mathbf{x}_t$ and count the frequency of each label. We treat the label with the largest frequency as the predicted label $\hat{y}_t$. Recall that, based on the definition of label probability, a global model trained on a random subsample with $k$ clients  predicts label $\hat{y}_t$ for  $\mathbf{x}_t$ with the label probability $p_{\hat{y}_t}$. Therefore, the frequency $N_{\hat{y}_t}$ of the label $\hat{y}_t$ among the $N$ global models follows a binomial distribution $B(N, p_{\hat{y}_t})$ with parameters $N$ and $p_{\hat{y}_t}$. Thus, given $N_{\hat{y}_t}$ and $N$, we can use the standard one-sided Clopper-Pearson method~\cite{clopper1934use} to estimate a lower bound $\underline{p_{\hat{y}_t}}$ of ${p_{\hat{y}_t}}$ with a confidence level $1-\alpha$. Specifically, we have $\underline{p_{\hat{y}_t}}=\mathcal{B}\left({\alpha};N_{\hat{y}_t},N-N_{\hat{y}_t}+1\right)$, where $\mathcal{B}(q; v,w)$ is the $q$th quantile from a beta distribution with shape parameters $v$ and $w$. 
Moreover, we can estimate $\overline{p}_{\hat{z}_t}=1-\underline{p_{\hat{y}_t}}\geq 1-p_{\hat{y}_t}\geq {p}_{z_t}$ as an upper bound of ${p}_{\hat{z}_t}$. 

\myparatight{Computing predicted labels and probability bounds for $d$ testing examples} One  method to compute the predicted labels and probability bounds for the $d$ testing examples is to apply the above process to each testing example individually. However, such method is computationally intractable because it requires training $N$ global models for every testing example. To address the computational challenge, we propose a method that only needs to train $N$ global models in total.  Our idea is to split $\alpha$ among the $d$ testing examples. Specifically, we follow the above process to train $N$ global models and use them to predict labels for the $d$ testing examples. For each testing example $\mathbf{x}_t$, we estimate the lower bound $\underline{p_{\hat{y}_t}}=\mathcal{B}\left(\frac{\alpha}{d};N_{\hat{y}_t},N-N_{\hat{y}_t}+1\right)$ with confidence level $1-\alpha/d$ instead of $1-\alpha$.  According to the \emph{Bonferroni correction}, 
  the simultaneous confidence level of estimating the lower bounds for the $d$ testing examples is $1-\alpha$. Following the above process, we still estimate $\overline{p}_{\hat{z}_t}=1-\underline{p_{\hat{y}_t}}$ as an upper bound of ${p}_{\hat{z}_t}$ for each testing example. 

\begin{algorithm}[tb]
   \caption{Computing Predicted Label and Certified Security Level}
   \label{alg:certify}
\begin{algorithmic}[1]
   \STATE {\bfseries Input:} $\mathbf{C}$, $\mathcal{A}$, $k$, $N$, $\mathcal{D}$, $\alpha$.
   \STATE {\bfseries Output:} Predicted label and certified security level for each testing example in $\mathcal{D}$. \\
   $g_1,g_2,\cdots,g_N \gets  \textsc{Sample\&Train}(\mathbf{C},\mathcal{A},k,N)$ \\
   \FOR{$\mathbf{x}_t$ {\bfseries in} $\mathcal{D}$}
   \STATE counts$[i] \gets \sum_{l=1}^{N}\mathbb{I}(g_{l}(\mathbf{x}_t)=i), i\in \{1,2,\cdots,L\} $   
   \STATE /* $\mathbb{I}$ is the indicator function */\\ 
   \STATE $\hat{y}_t\gets$  index of the largest entry in counts (ties are broken uniformly at random) \\
   \STATE $\underline{p_{\hat{y}_t}}\gets \mathcal{B}\left(\frac{\alpha}{d};N_{\hat{y}_t},N-N_{\hat{y}_t}+1\right)$ \\
   \STATE $ \overline{p}_{\hat{z}_t} \gets 1-\underline{p_{\hat{y}_t}}$
   \IF{$\underline{p_{\hat{y}_t}} > \overline{p}_{\hat{z}_t}$}
   \STATE $\hat{m}_t^* \gets \textsc{SearchLevel} (\underline{p_{\hat{y}_t}}, \overline{p}_{\hat{z}_t}, k, |\mathbf{C}|)$ \\
   \ELSE
   \STATE $\hat{y}_t \gets \text{ABSTAIN}$, $\hat{m}_t^* \gets \text{ABSTAIN}$
   \ENDIF
   \ENDFOR
  \STATE \textbf{return} $\hat{y}_1,\hat{y}_2,\cdots, \hat{y}_d$ and $\hat{m}_1^*,\hat{m}_2^*,\cdots, \hat{m}_d^*$
\end{algorithmic}
\end{algorithm}

\noindent
{\bf Complete algorithm:}
Algorithm~\ref{alg:certify} shows our algorithm to compute the predicted labels and certified security levels for the $d$ testing examples in $\mathcal{D}$. The function \textsc{Sample\&Train} randomly samples $N$ subsamples with $k$ clients and trains $N$ global models using the base federated learning algorithm $\mathcal{A}$. Given the probability bounds $\underline{p_{\hat{y}_t}}$ and $\overline{p}_{\hat{z}_t}$ for a testing example $\mathbf{x}_t$, the function \textsc{SearchLevel} finds the certified security level $\hat{m}_t^*$  via finding the largest integer $m$ that satisfies (\ref{eq:certified_condition1}).  For example, \textsc{SearchLevel} can simply start $m$ from 0 and iteratively increase it by one until finding $\hat{m}_t^*$. 

\noindent
{\bf Probabilistic guarantees:}   In Algorithm~\ref{alg:certify}, since we estimate the lower bound $\underline{p_{\hat{y}_t}}$ using the Clopper-Pearson method, there is a probability that the estimated lower bound is incorrect, i.e., $\underline{p_{\hat{y}_t}} > p_{\hat{y}_t}$. When the lower bound is estimated incorrectly for a testing example $\mathbf{x}_t$, the certified security level $\hat{m}_t^*$ outputted by Algorithm~\ref{alg:certify} for $\mathbf{x}_t$ may also be incorrect, i.e., there may exist an $\mathbf{C}'$ such that $M(\mathbf{C}')\leq \hat{m}_t^*$ but $h(\mathbf{C}',\mathbf{x}_t)\neq \hat{y}_t$. In other words, our Algorithm~\ref{alg:certify} has probabilistic guarantees for its outputted certified security levels. However, in the following theorem, we prove the probability that Algorithm~\ref{alg:certify}  returns an incorrect certified security level for at least one testing example is at most $\alpha$. 
\begin{theorem}
\label{probability_of_certify}
The probability that Algorithm~\ref{alg:certify} returns an incorrect certified security level for at least one testing example in $\mathcal{D}$ is bounded by $\alpha$, which is equivalent to: 
\begin{align}
     \text{Pr}&(\cap_{\mathbf{x}_t \in \mathcal{D}} (h(\mathbf{C}',\mathbf{x}_t)=\hat{y}_t, \forall \mathbf{C}', M(\mathbf{C}')\leq \hat{m}_t^*|\hat{y}_t\neq \text{ABSTAIN})) \nonumber\\&\geq 1 -\alpha.
\end{align}
\end{theorem}

Note that when the probability bounds are estimated deterministically, e.g., when ${n \choose k}$ is small and the exact label probabilities can be computed  via training ${n \choose k}$ global models, the certified security level obtained from our Theorem~\ref{certified_radius} is also deterministic. 


\section{Experiments}

\begin{table}[!tb]
\centering
\scalebox{0.92}{
\begin{tabular} {|c|c|c|c|}\hline 
{\small Dataset} & {\small MNIST} & {\small HAR}\\ \hline
{\small Model architecture} & {\small CNN} & {\small DNN}\\\hline
{\small Number of clients} & {\small 1,000} & {\small 30}\\\hline
{\small $globalIter$} & {\small  3,000} & {\small 5,000}\\\hline
\makecell{\small $localIter$}  & \multicolumn{2}{c|}{\small 5}\\\hline
{\small Learning rate $\eta$} & \multicolumn{2}{c|}{\small 0.001}\\\hline
{\small Batch size} & \multicolumn{2}{c|}{\small 32}\\\hline
\end{tabular}
}
\caption{Federated learning settings and hyperparameters.}
\label{tab:hyperparameters}
\end{table}

\subsection{Experimental Setup}

\noindent
{\bf Datasets, model architectures, and base algorithm:} We use MNIST \cite{lecun2010mnist} and Human Activity Recognition (HAR) datasets \cite{anguita2013public}.  MNIST is used to simulate federated learning scenarios, while  HAR represents a real-world federated learning scenario. Specifically, MNIST has 60,000 training examples and 10,000 testing examples. We consider $n=1,000$ clients and we split them into $10$ groups. We assign a training example with label $l$ to the $l$th group of clients with probability $q$ and assign it to each remaining group with a probability $\frac{1-q}{9}$. After assigning a training example to a group, we distribute it to a client in the group uniformly at random. 
The parameter $q$ controls local training data distribution on clients and we call $q$ \emph{degree of non-IID}.  $q=0.1$ means that clients' local training data are IID, while a larger $q$ indicates a larger {degree of non-IID}. By default, we set $q=0.5$. However, we will study the impact of $q$ (degree of non-IID) on our method. HAR includes human activity data from 30 users, each of which is a client. The task is to predict a user's activity based on the sensor signals (e.g., acceleration) collected from the user's smartphone.  There are 6 possible activities (e.g., walking, sitting, and standing), indicating  a 6-class classification problem.  There are 10,299 examples in total and each example has 561 features. We use 75\% of each user's examples as training examples and the rest as testing examples.  

We consider a convolutional neural network (CNN) architecture (shown in Supplemental Material) for MNIST. For HAR, we consider a deep neural network (DNN) with two fully-connected hidden layers, each of which contains 256 neurons and uses ReLU as the activation function. We use the popular FedAvg~\cite{mcmahan2016communication} as the base federated learning algorithm. Recall that a base federated learning algorithm has hyperparameters (shown in Algorithm~\ref{alg:base_alg}): $globalIter$, $localIter$,  learning rate $\eta$, and batch size. Table \ref{tab:hyperparameters} summarizes these hyperparameters for FedAvg in our experiments. In particular, we set the $globalIter$ in Table \ref{tab:hyperparameters} because FedAvg converges with such settings.

\begin{figure}[!t]
    \center
    \vspace{-2mm}
    \subfigure[MNIST]{\includegraphics[width=0.23\textwidth]{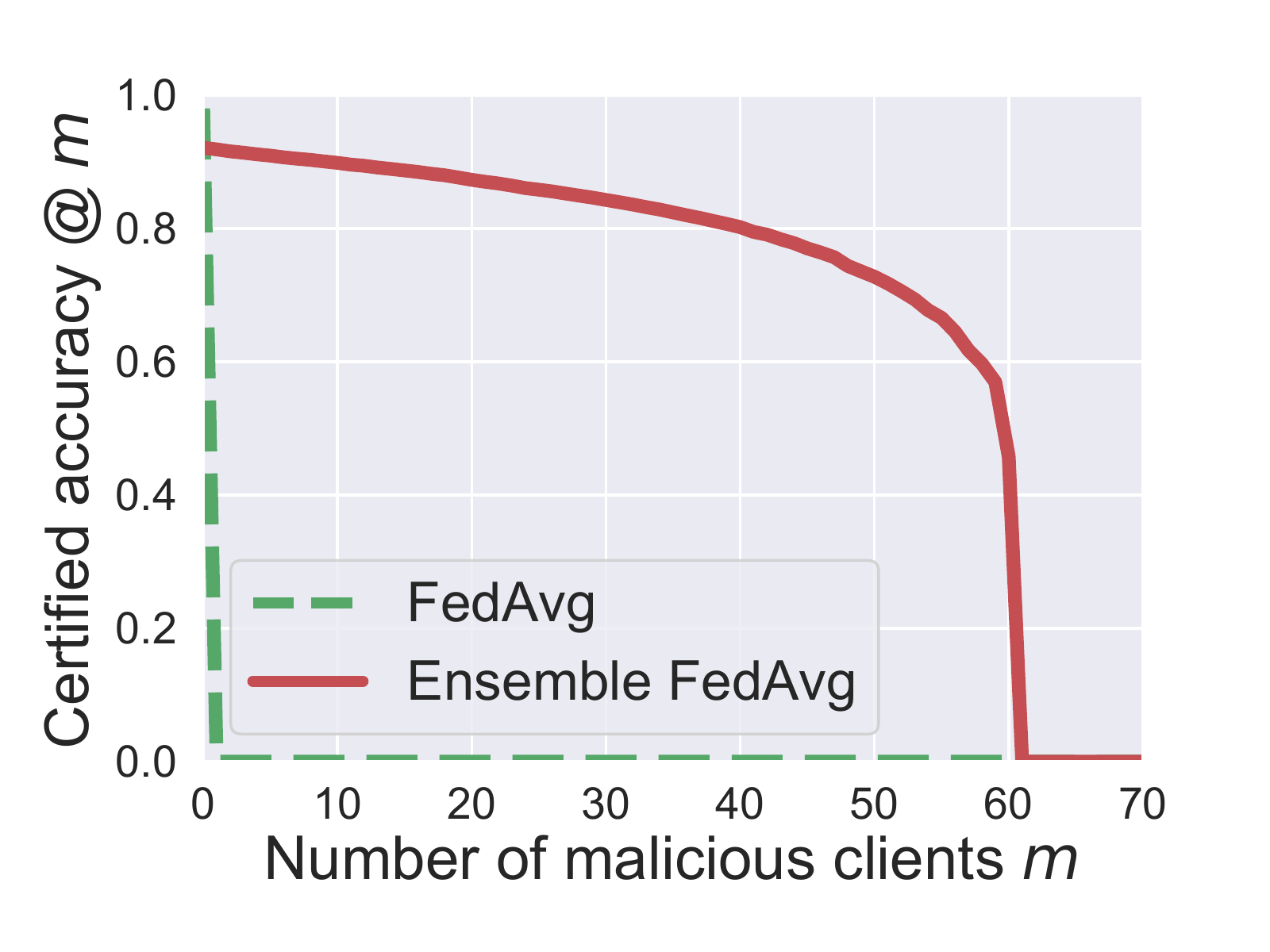}\label{fig:mnist_compare}}
    \subfigure[HAR]{\includegraphics[width=0.23\textwidth]{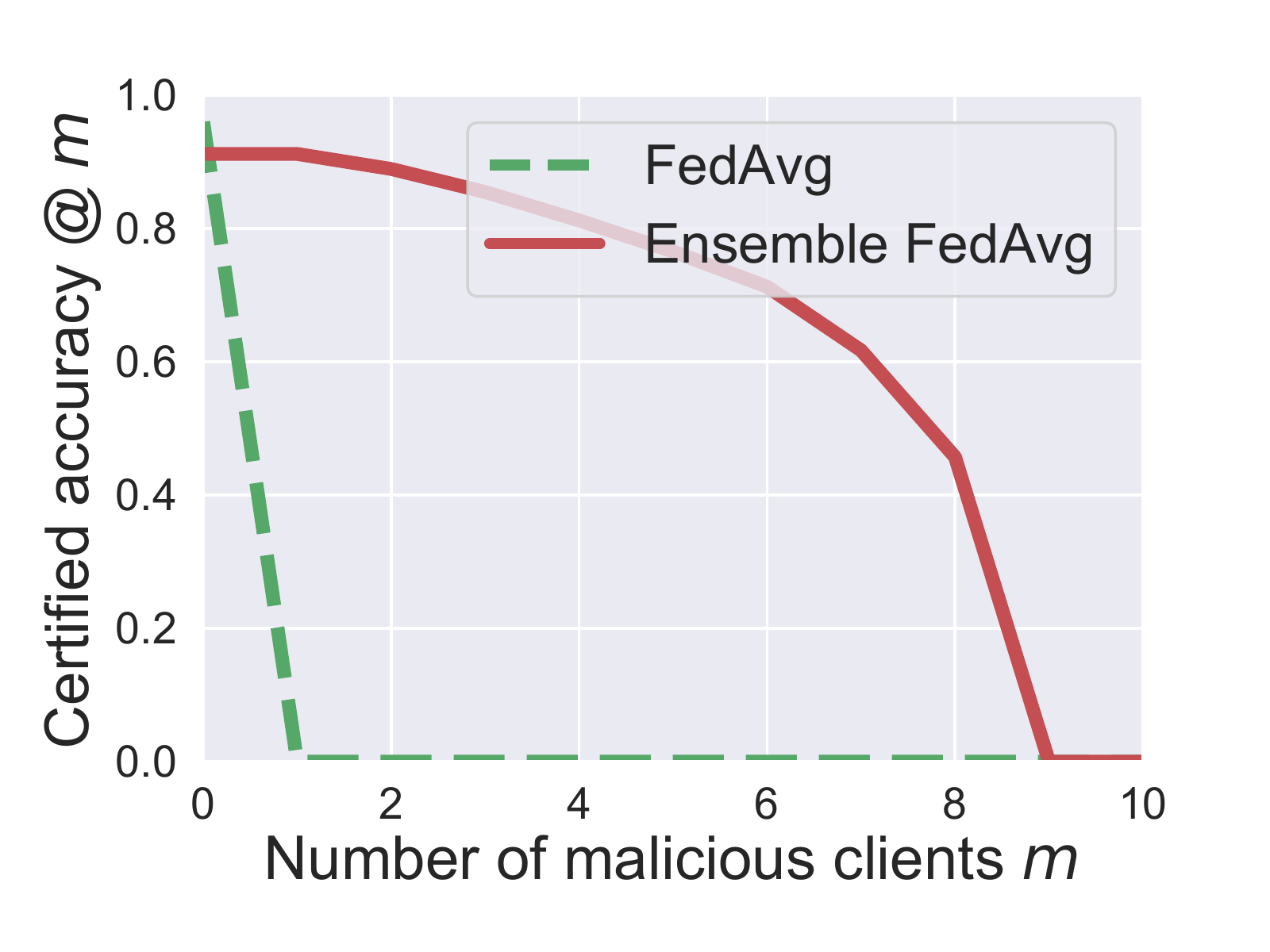}\label{fig:human_compare}}
    \vspace{-3mm}
    \caption{FedAvg vs. ensemble FedAvg.}
    \label{fig:q_compare}
\vspace{-3mm}
\end{figure}

\begin{figure}[!t]
    \center
    \subfigure[MNIST]{\includegraphics[width=0.23\textwidth]{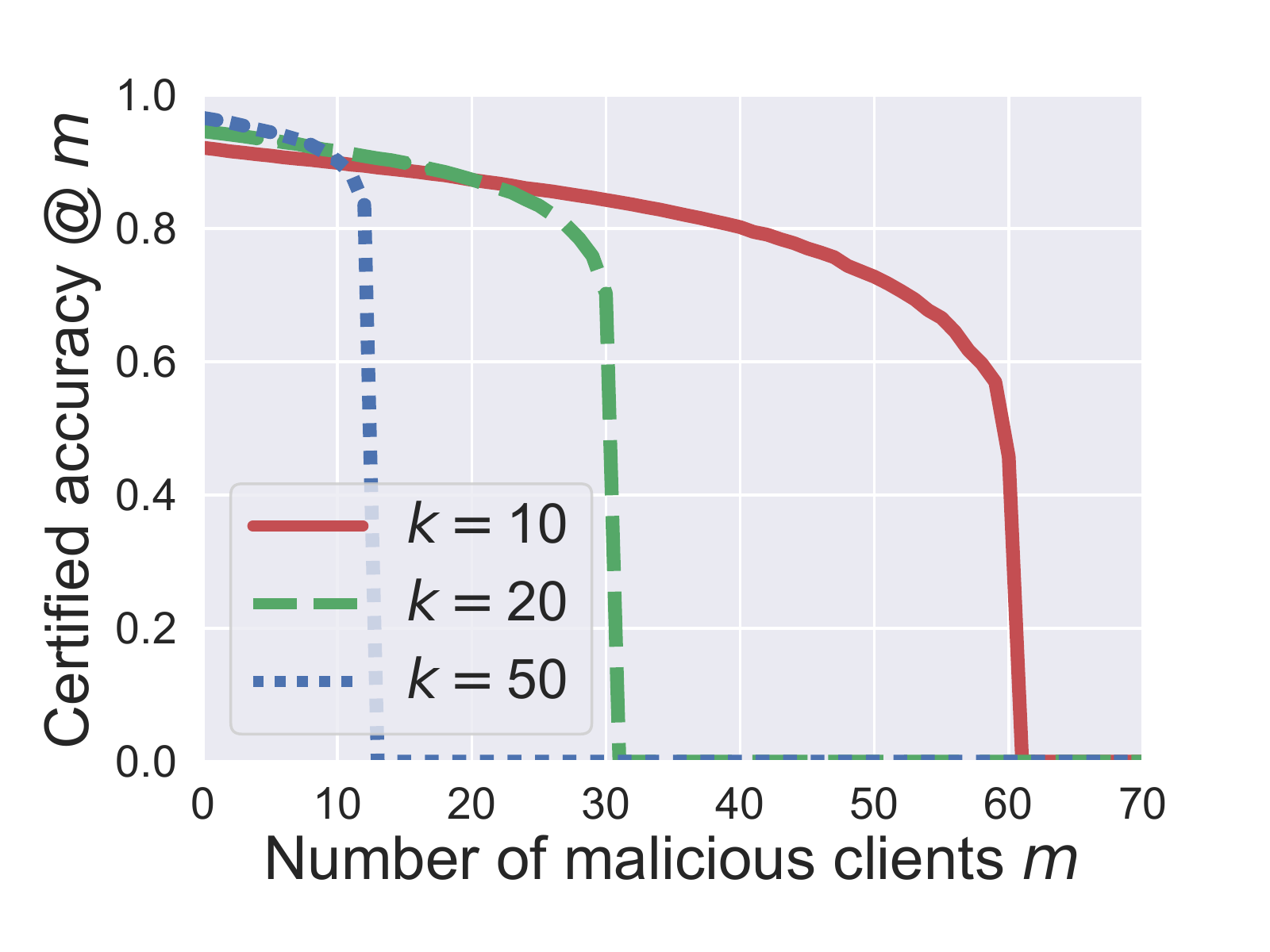}}
    \subfigure[HAR]{\includegraphics[width=0.23\textwidth]{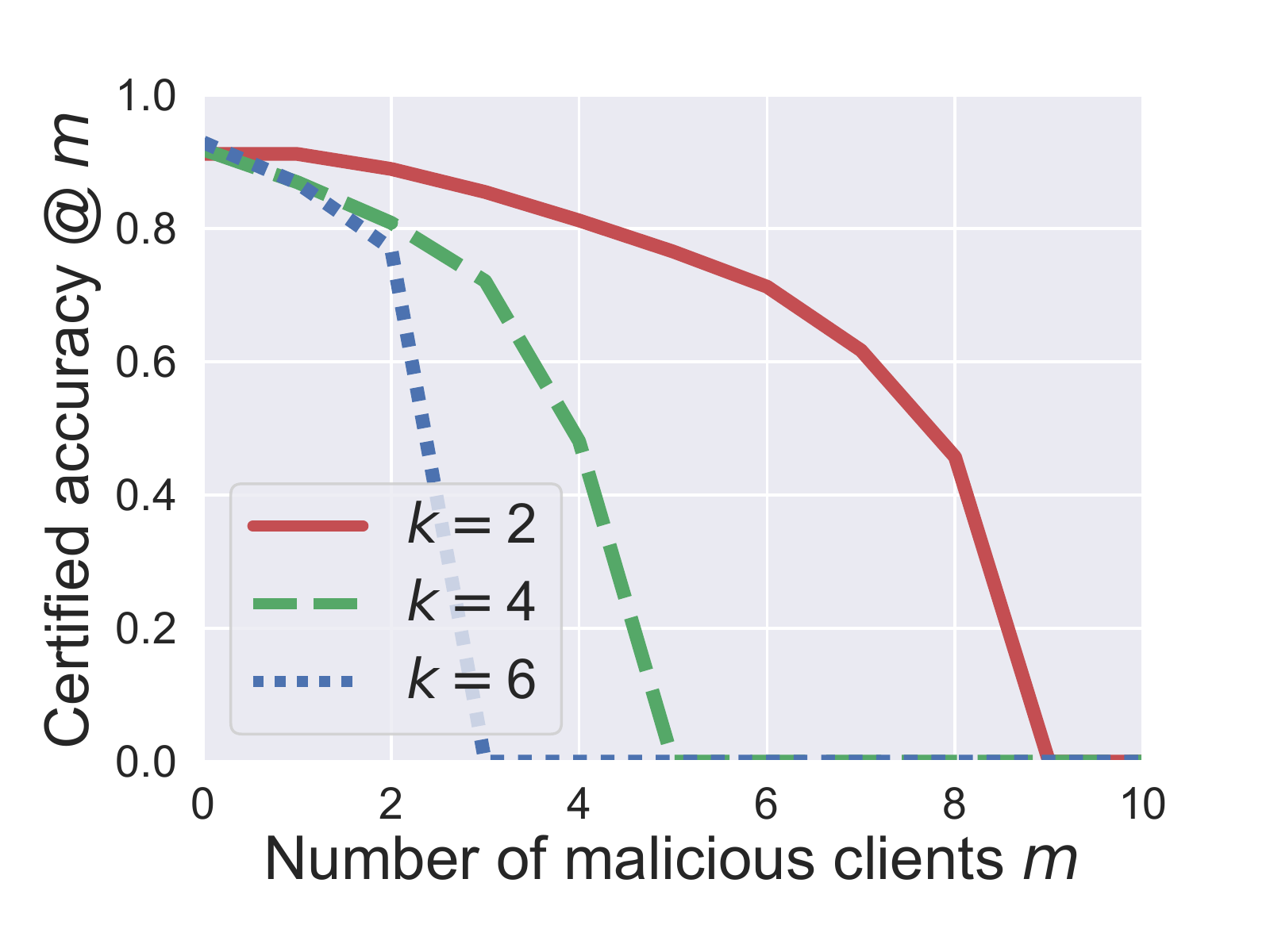}\label{fig:k-Human}}  
    \vspace{-3mm}
    \caption{Impact of $k$ on our ensemble FedAvg.}
    \label{fig:k}
\vspace{-1mm}
\end{figure}

\noindent
{\bf Evaluation metric:} 
We use  \emph{certified accuracy} as our evaluation metric. 
Specifically, we define the \emph{certified accuracy at $m$ malicious clients} (denoted as CA$@m$) for a federated learning method as the fraction of testing examples in the testing dataset $\mathcal{D}$ whose labels are correctly predicted by the method and whose certified security levels are at least $m$. Formally, we define CA$@m$ as follows: 
\begin{align}
\text{CA}@m=\frac{\sum_{\mathbf{x}_t \in \mathcal{D}} \mathbb{I}(\hat{y}_t=y_t)\cdot \mathbb{I}(\hat{m}_{t}^{*}\geq m)}{|\mathcal{D}|},
\end{align}
where $\mathbb{I}$ is the indicator function, $y_t$ is the true label for $\mathbf{x}_t$, and $\hat{y}_t$ and $\hat{m}_{t}^{*}$ are respectively the predicted label and  certified security level for $\mathbf{x}_t$. Intuitively, CA$@m$ means that when at most $m$ clients are malicious, the accuracy of the federated learning method for $\mathcal{D}$ is at least CA$@m$ no matter what attacks the malicious clients use (i.e., no matter how the malicious clients tamper their local training data and model updates). Note that CA$@0$ reduces to the standard accuracy when there are no malicious clients. 

When we can compute the exact label probabilities via training ${n \choose k}$ global models, the CA$@m$ of our ensemble global model $h$ computed using the certified security levels derived from Theorem~\ref{certified_radius} is deterministic. When ${n \choose k}$ is large, we estimate predicted labels and certified security levels using Algorithm~\ref{alg:certify}, and thus our CA$@m$ has a confidence level $1-\alpha$ according to Theorem~\ref{probability_of_certify}. 

\noindent
{\bf Parameter settings:}  Our method has three parameters: $N$, $k$, and $\alpha$. Unless otherwise mentioned, we adopt the following default settings for them: $N=500$, $\alpha=0.001$,  $k=10$ for MNIST, and $k=2$ for HAR.  
Under such default setting for HAR, we have ${n \choose k}={30 \choose 2} = 435 < N=500$ and  we can compute the exact label probabilities via training 435 global models. Therefore, we have deterministic certified accuracy for HAR under the default setting. We will explore the impact of each parameter while using the default settings for the other two parameters. For HAR, we set $k=4$ when exploring the impact of $N$ (i.e., Figure~\ref{impactNh}) and $\alpha$ (i.e., Figure~\ref{impactah}) since the default setting $k=2$ gives deterministic certified accuracy, making $N$ and $\alpha$ not relevant. 

\subsection{Experimental Results}

\begin{figure}[!t]
    \center
    \vspace{-2mm}
     \subfigure[MNIST]{\includegraphics[width=0.23\textwidth]{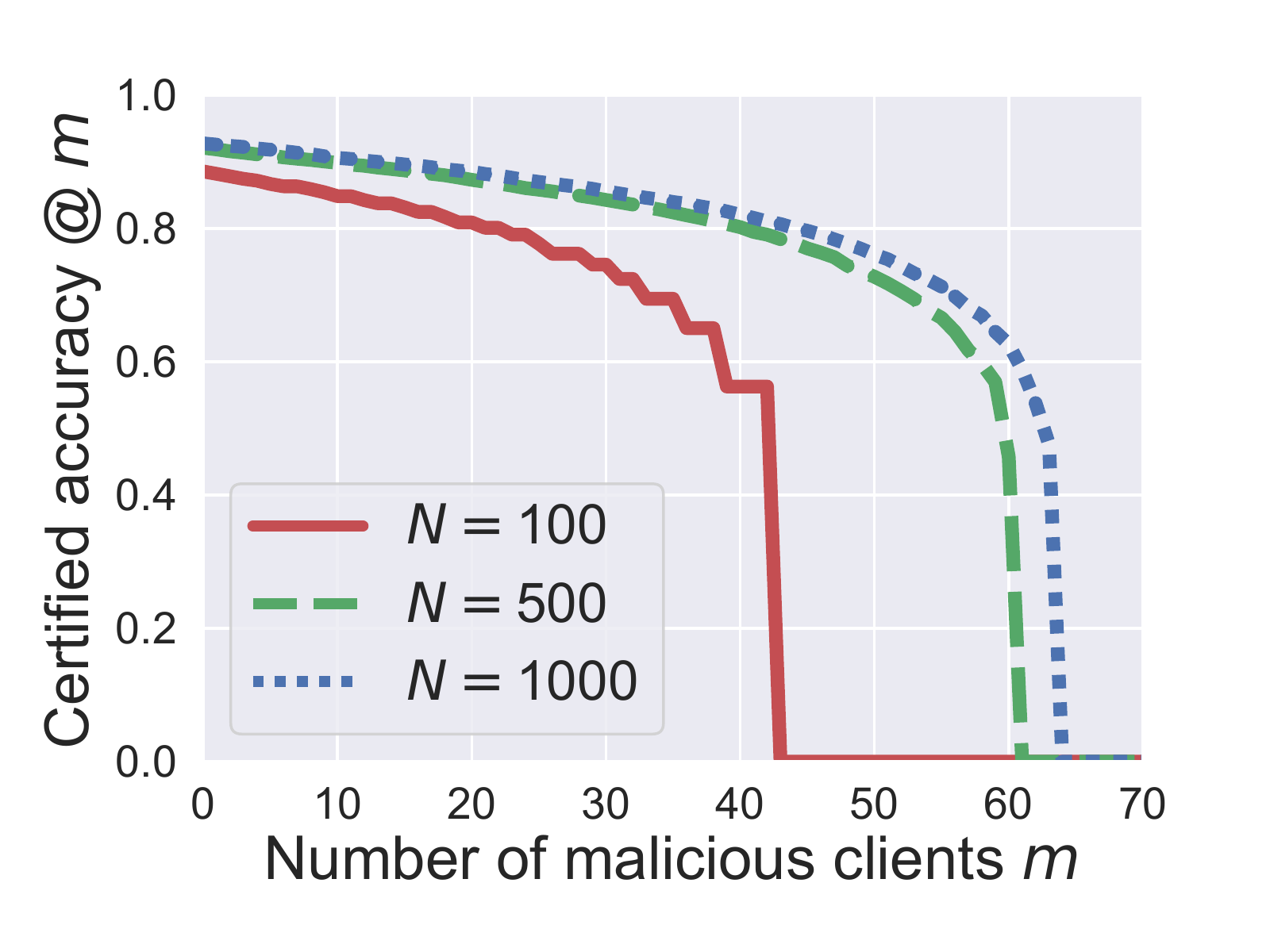}}
    \subfigure[HAR]{\includegraphics[width=0.23\textwidth]{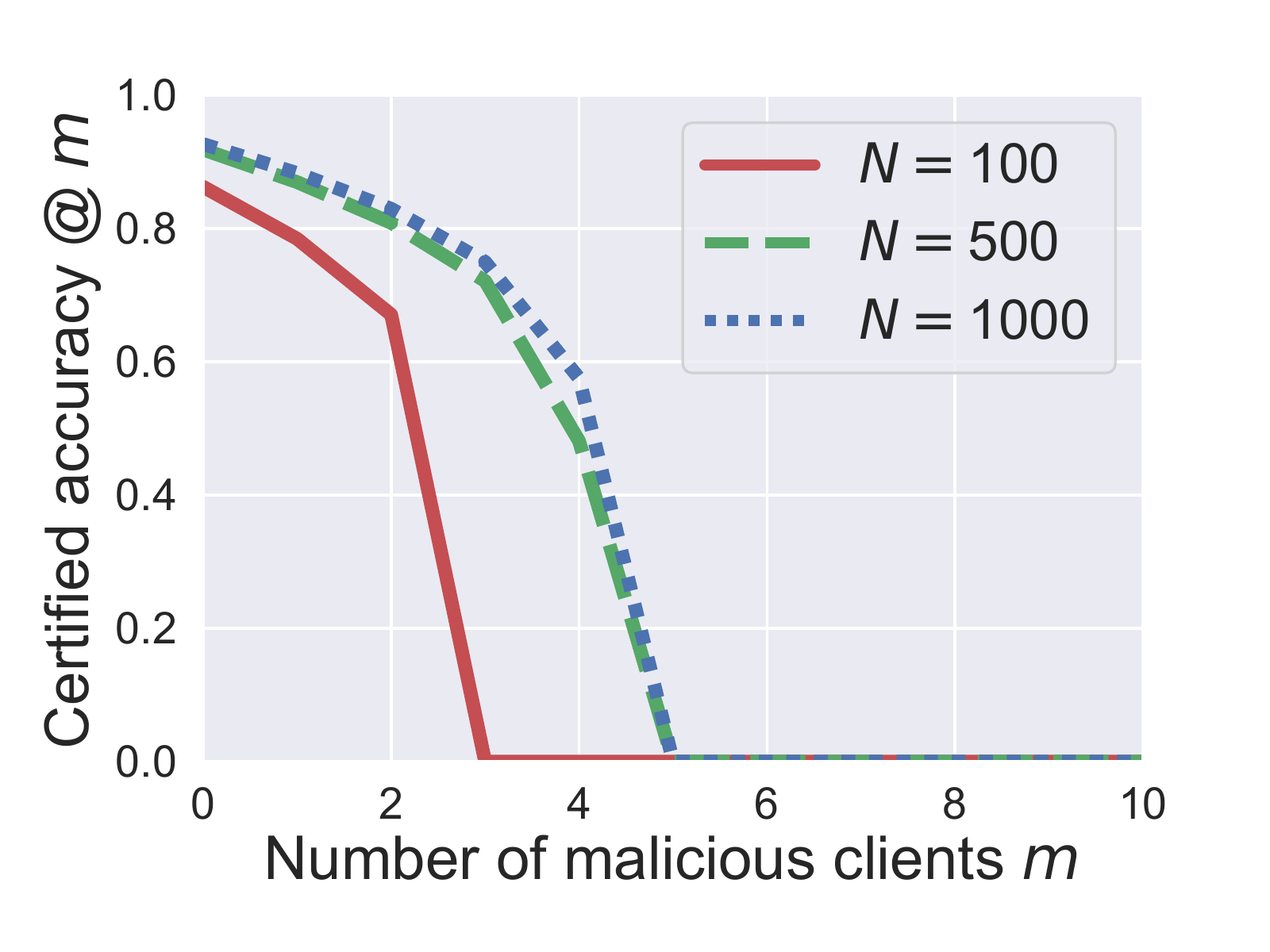}\label{impactNh}}
    \vspace{-3mm}
    \caption{Impact of $N$ on our ensemble FedAvg.}
    \label{fig:N}
\vspace{-3mm}
\end{figure}

\begin{figure}[!t]
    \center
     \subfigure[MNIST]{\includegraphics[width=0.23\textwidth]{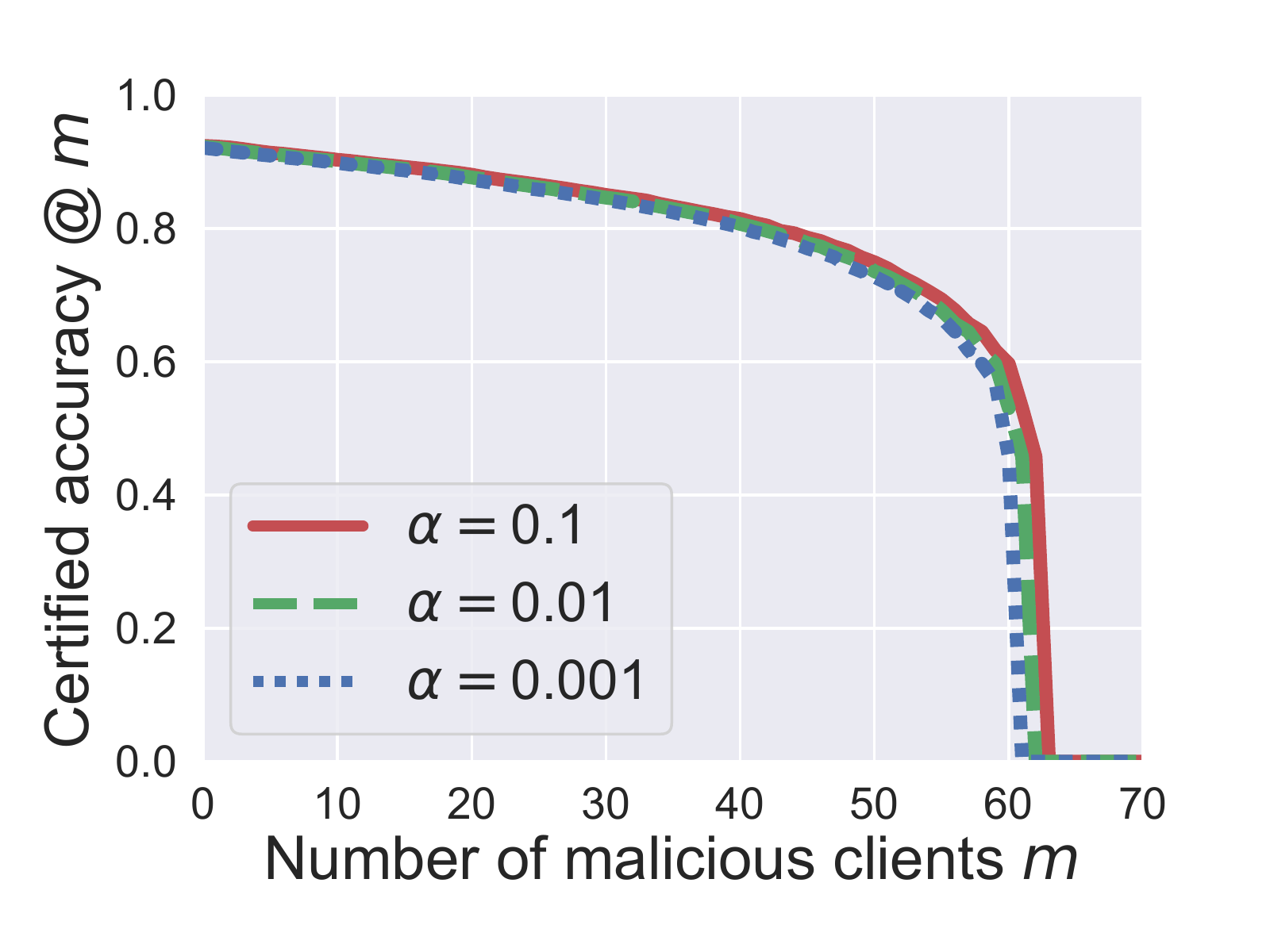}} 
    \subfigure[HAR]{\includegraphics[width=0.23\textwidth]{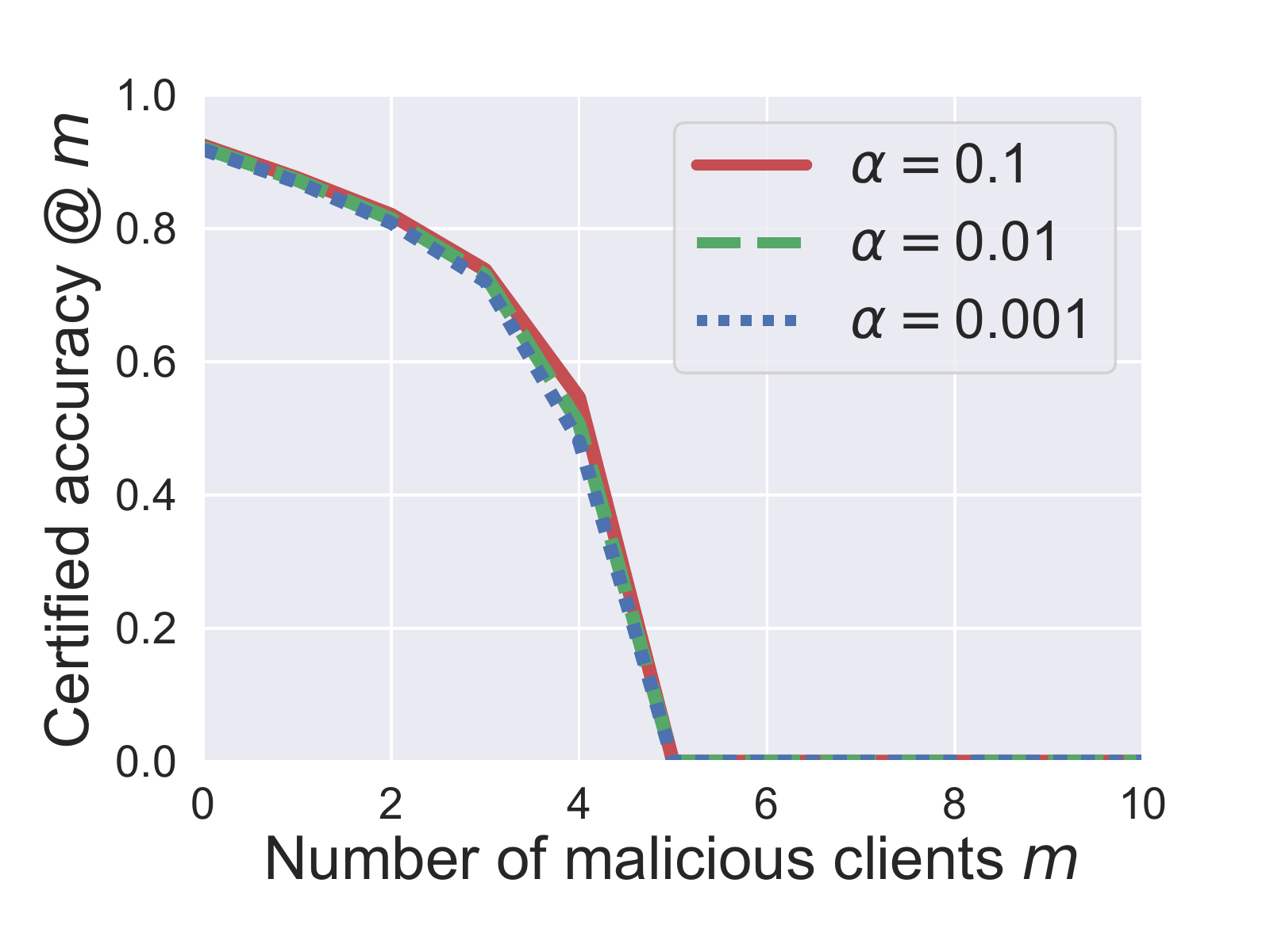}\label{impactah}}
    \vspace{-3mm}
    \caption{Impact of $\alpha$ on our ensemble FedAvg.}
    \label{fig:alpha}
\vspace{-1mm}
\end{figure}

\noindent
{\bf Single-global-model FedAvg vs. ensemble FedAvg:} Figure~\ref{fig:q_compare} compares single-global-model FedAvg and ensemble FedAvg with respect to certified accuracy on the two datasets. 
When there are no malicious clients (i.e., $m=0$), single-global-model FedAvg is more accurate than ensemble FedAvg. This is because ensemble FedAvg uses a subsample of clients to train each global model. 
However, single-global-model FedAvg has 0 certified accuracy when just one client is malicious. This is because a single malicious client can arbitrarily manipulate the global model learnt by FedAvg~\cite{Blanchard17}. However, the certified accuracy of ensemble FedAvg reduces to 0 when up to 61 and 9 clients (6.1\% and 30\%) are malicious on MNIST and HAR, respectively. Note that it is unknown whether existing Byzantine-robust federated learning methods have non-zero certified accuracy when $m>0$, and thus we cannot compare ensemble FedAvg with them.  

\myparatight{Impact of $k$, $N$, and $\alpha$} Figure \ref{fig:k}, \ref{fig:N}, and  \ref{fig:alpha} show the impact of $k$, $N$, and $\alpha$, respectively. $k$ achieves a trade-off between accuracy under no malicious clients and security under malicious clients. 
Specifically, when $k$ is larger, the ensemble global model is more accurate at $m=0$, but the certified accuracy drops more quickly to 0 as $m$ increases. This is because when $k$ is larger, it is more likely for the sampled $k$ clients to include malicious ones.  
The certified accuracy increases as $N$ or $\alpha$ increases. This is because training more global models or a larger $\alpha$ allows Algorithm~\ref{alg:certify} to estimate tighter probability bounds and larger certified security levels. When $N$ increases from 100 to 500,  the certified accuracy increases significantly. However, when $N$ further grows to 1,000, the increase of certified accuracy is marginal. Our results show that  we don't need to train too many global models in practice, as the certified accuracy saturates when $N$ is larger than some threshold.   

\begin{figure}[!t]
    \center
    \includegraphics[width=0.23\textwidth]{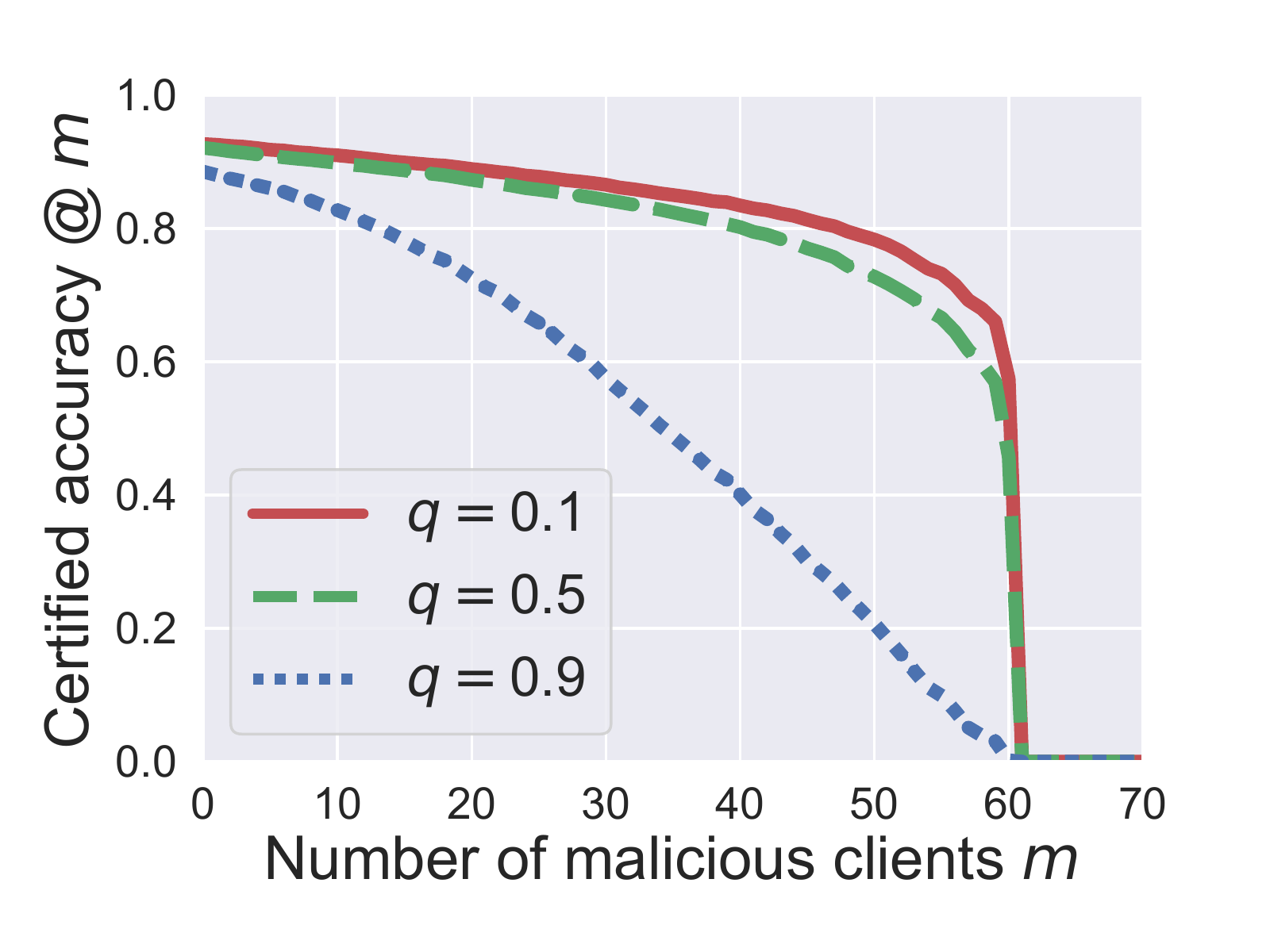}
    \vspace{-3mm}
    \caption{Impact of the degree of non-IID $q$ on MNIST.}
    \label{fig:q}
\end{figure}

\myparatight{Impact of degree of non-IID $q$} Figure~\ref{fig:q} shows the certified accuracy of our ensemble FedAvg on MNIST when the clients' local training data have different degrees of non-IID. We observe that the certified accuracy drops when $q$ increases from 0.5 to 0.9, which represents a high degree of non-IID. However, the certified accuracy is still high when $m$ is small for $q=0.9$, e.g., the certified accuracy is still 83\% when $m=10$.  This is because  although each global model trained using a subsample of clients is less accurate when the local training data are highly non-IID, the ensemble of multiple global models is still accurate.


\section{Related Work}
In federated learning, the first category of studies~\cite{smith2017federated,li2019convergence,wang2020federated,liu2020federated,peng2019federated} aim to design  federated learning methods that can learn more accurate global models and/or analyze their convergence properties. For instance, FedMA \cite{wang2020federated} constructs the global model via matching and averaging the hidden elements in a neural network with similar feature extraction signatures. The second category of studies~\cite{konevcny2016federated,mcmahan2016communication,wen2017terngrad,alistarh2017qsgd,lee2017speeding,sahu2018convergence,bernstein2018signsgd,vogels2019powersgd,yurochkin2019bayesian,mohri2019agnostic,wang2020federated,li2020practical,li2020acceleration,hamer2020fedboost,rothchildfetchsgd,malinovsky2020local} aim to improve the communication efficiency between the clients and server via sparsification, quantization, and/or encoding of the model updates sent from the clients to the server. The third category of studies~\cite{bonawitz2017practical,geyer2017differentially,hitaj2017deep,melis2019exploiting,zhu2019deep,mohri2019agnostic,wang2020federatedlatent,Li2020Fair} aim to explore the privacy/fairness issues of federated learning and their defenses.These studies often assume a single global model is shared among the clients.  Smith et al. \cite{smith2017federated} proposed to learn a customized model for each client via multi-task learning.

Our work is on  security of federated learning, which is orthogonal to the studies above. Multiple studies~\cite{fang2019local,Bagdasaryan18,Xie19,Bhagoji19} showed that the  global model's accuracy can be significantly downgraded by malicious clients. Existing defenses against malicious clients leverage Byzantine-robust aggregation rules such as Krum~\cite{Blanchard17}, trimmed mean \cite{Yin18},  coordinate-wise median \cite{Yin18}, and Bulyan~\cite{Mhamdi18}. However, they cannot provably guarantee that the global model's predicted label for a testing example is not affected by malicious clients. As a result, they may be broken by strong attacks that carefully craft the model updates sent from the malicious clients to the server, e.g., \cite{fang2019local}. We propose ensemble federated learning whose predicted label for a testing example is provably not affected by a bounded number of malicious clients. 

We note that ensemble methods were also proposed as provably secure defenses (e.g.,~\cite{jia2020intrinsic}) against data poisoning attacks. However, they are insufficient to defend against malicious clients that can manipulate both the local training data and the model updates. In particular, a provably secure defense against data poisoning attacks guarantees that the label predicted for a testing example is unaffected by a bounded number of poisoned training examples. However, a single malicious client can poison an arbitrary number of its local training examples, breaking the assumption of provably secure defenses against data poisoning attacks. 


\section{Conclusion}
In this work, we propose ensemble federated learning and derive its tight provable security guarantee against malicious clients. Moreover, we propose an algorithm to compute the certified security levels. Our empirical results on two datasets show that our ensemble federated learning can effectively defend against malicious clients with provable security guarantees.  Interesting future work includes estimating the probability bounds deterministically and considering the internal structure of a base federated learning algorithm to further improve our provable security guarantees. 


\section*{Acknowledgement}
We thank the anonymous reviewers for insightful reviews. This work was supported by NSF grant No.1937786.


\bibliography{refs}

\begin{thebibliography}{44}
\providecommand{\natexlab}[1]{#1}
\providecommand{\url}[1]{\texttt{#1}}
\providecommand{\urlprefix}{URL }
\expandafter\ifx\csname urlstyle\endcsname\relax
  \providecommand{\doi}[1]{doi:\discretionary{}{}{}#1}\else
  \providecommand{\doi}{doi:\discretionary{}{}{}\begingroup
  \urlstyle{rm}\Url}\fi

\bibitem[{Alistarh, Allen-Zhu, and Li(2018)}]{alistarh2018byzantine}
Alistarh, D.; Allen-Zhu, Z.; and Li, J. 2018.
\newblock Byzantine stochastic gradient descent.
\newblock In \emph{NeurIPS}.

\bibitem[{Alistarh et~al.(2017)Alistarh, Grubic, Li, Tomioka, and
  Vojnovic}]{alistarh2017qsgd}
Alistarh, D.; Grubic, D.; Li, J.; Tomioka, R.; and Vojnovic, M. 2017.
\newblock QSGD: Communication-efficient SGD via gradient quantization and
  encoding.
\newblock In \emph{NeurIPS}.

\bibitem[{Anguita et~al.(2013)Anguita, Ghio, Oneto, Parra, and
  Reyes-Ortiz}]{anguita2013public}
Anguita, D.; Ghio, A.; Oneto, L.; Parra, X.; and Reyes-Ortiz, J.~L. 2013.
\newblock A public domain dataset for human activity recognition using
  smartphones.
\newblock In \emph{ESANN}.

\bibitem[{Bagdasaryan et~al.(2020)Bagdasaryan, Veit, Hua, Estrin, and
  Shmatikov}]{Bagdasaryan18}
Bagdasaryan, E.; Veit, A.; Hua, Y.; Estrin, D.; and Shmatikov, V. 2020.
\newblock How to backdoor federated learning.
\newblock In \emph{AISTATS}.

\bibitem[{Bernstein et~al.(2018)Bernstein, Wang, Azizzadenesheli, and
  Anandkumar}]{bernstein2018signsgd}
Bernstein, J.; Wang, Y.-X.; Azizzadenesheli, K.; and Anandkumar, A. 2018.
\newblock signSGD: Compressed Optimisation for Non-Convex Problems.
\newblock In \emph{ICML}.

\bibitem[{Bhagoji et~al.(2019)Bhagoji, Chakraborty, Mittal, and
  Calo}]{Bhagoji19}
Bhagoji, A.; Chakraborty, S.; Mittal, P.; and Calo, S. 2019.
\newblock Analyzing Federated Learning through an Adversarial Lens.
\newblock In \emph{ICML}.

\bibitem[{Blanchard et~al.(2017)Blanchard, Mhamdi, Guerraoui, and
  Stainer}]{Blanchard17}
Blanchard, P.; Mhamdi, E. M.~E.; Guerraoui, R.; and Stainer, J. 2017.
\newblock Machine Learning with Adversaries: Byzantine Tolerant Gradient
  Descent.
\newblock In \emph{NeurIPS}.

\bibitem[{Bonawitz et~al.(2017)Bonawitz, Ivanov, Kreuter, Marcedone, McMahan,
  Patel, Ramage, Segal, and Seth}]{bonawitz2017practical}
Bonawitz, K.; Ivanov, V.; Kreuter, B.; Marcedone, A.; McMahan, H.~B.; Patel,
  S.; Ramage, D.; Segal, A.; and Seth, K. 2017.
\newblock Practical secure aggregation for privacy-preserving machine learning.
\newblock In \emph{CCS}.

\bibitem[{Chen et~al.(2018)Chen, Wang, Charles, and
  Papailiopoulos}]{chen2018draco}
Chen, L.; Wang, H.; Charles, Z.; and Papailiopoulos, D. 2018.
\newblock DRACO: Byzantine-resilient Distributed Training via Redundant
  Gradients.
\newblock In \emph{ICML}.

\bibitem[{Chen, Su, and Xu(2017)}]{ChenPOMACS17}
Chen, Y.; Su, L.; and Xu, J. 2017.
\newblock Distributed Statistical Machine Learning in Adversarial Settings:
  Byzantine Gradient Descent.
\newblock In \emph{POMACS}.

\bibitem[{Clopper and Pearson(1934)}]{clopper1934use}
Clopper, C.~J.; and Pearson, E.~S. 1934.
\newblock The use of confidence or fiducial limits illustrated in the case of
  the binomial.
\newblock \emph{Biometrika} .

\bibitem[{Fang et~al.(2020)Fang, Cao, Jia, and Gong}]{fang2019local}
Fang, M.; Cao, X.; Jia, J.; and Gong, N.~Z. 2020.
\newblock Local model poisoning attacks to Byzantine-robust federated learning.
\newblock In \emph{USENIX Security}.

\bibitem[{Geyer, Klein, and Nabi(2017)}]{geyer2017differentially}
Geyer, R.~C.; Klein, T.; and Nabi, M. 2017.
\newblock Differentially private federated learning: A client level
  perspective.
\newblock \emph{arXiv preprint arXiv:1712.07557} .

\bibitem[{Hamer, Mohri, and Suresh(2020)}]{hamer2020fedboost}
Hamer, J.; Mohri, M.; and Suresh, A.~T. 2020.
\newblock FedBoost: Communication-Efficient Algorithms for Federated Learning.
\newblock In \emph{ICML}.

\bibitem[{Hitaj, Ateniese, and Perez-Cruz(2017)}]{hitaj2017deep}
Hitaj, B.; Ateniese, G.; and Perez-Cruz, F. 2017.
\newblock Deep models under the GAN: information leakage from collaborative
  deep learning.
\newblock In \emph{CCS}.

\bibitem[{Jia, Cao, and Gong(2020)}]{jia2020intrinsic}
Jia, J.; Cao, X.; and Gong, N.~Z. 2020.
\newblock Intrinsic certified robustness of bagging against data poisoning
  attacks.
\newblock \emph{arXiv preprint arXiv:2008.04495} .

\bibitem[{Kairouz et~al.(2019)Kairouz, McMahan, Avent, Bellet, Bennis, Bhagoji,
  Bonawitz, Charles, Cormode, Cummings et~al.}]{kairouz2019advances}
Kairouz, P.; McMahan, H.~B.; Avent, B.; Bellet, A.; Bennis, M.; Bhagoji, A.~N.;
  Bonawitz, K.; Charles, Z.; Cormode, G.; Cummings, R.; et~al. 2019.
\newblock Advances and open problems in federated learning.
\newblock \emph{arXiv preprint arXiv:1912.04977} .

\bibitem[{Kone{\v{c}}n{\`y} et~al.(2016)Kone{\v{c}}n{\`y}, McMahan, Yu,
  Richt{\'a}rik, Suresh, and Bacon}]{konevcny2016federated}
Kone{\v{c}}n{\`y}, J.; McMahan, H.~B.; Yu, F.~X.; Richt{\'a}rik, P.; Suresh,
  A.~T.; and Bacon, D. 2016.
\newblock Federated learning: Strategies for improving communication
  efficiency.
\newblock In \emph{NeurIPS Workshop on Private Multi-Party Machine Learning}.

\bibitem[{LeCun, Cortes, and Burges(1998)}]{lecun2010mnist}
LeCun, Y.; Cortes, C.; and Burges, C. 1998.
\newblock MNIST handwritten digit database.
\newblock \emph{Available: http://yann. lecun. com/exdb/mnist} .

\bibitem[{Lee et~al.(2017)Lee, Lam, Pedarsani, Papailiopoulos, and
  Ramchandran}]{lee2017speeding}
Lee, K.; Lam, M.; Pedarsani, R.; Papailiopoulos, D.; and Ramchandran, K. 2017.
\newblock Speeding up distributed machine learning using codes.
\newblock \emph{IEEE Transactions on Information Theory} .

\bibitem[{Li, Wen, and He(2020)}]{li2020practical}
Li, Q.; Wen, Z.; and He, B. 2020.
\newblock Practical Federated Gradient Boosting Decision Trees.
\newblock In \emph{AAAI}.

\bibitem[{Li et~al.(2020{\natexlab{a}})Li, Sanjabi, Beirami, and
  Smith}]{Li2020Fair}
Li, T.; Sanjabi, M.; Beirami, A.; and Smith, V. 2020{\natexlab{a}}.
\newblock Fair Resource Allocation in Federated Learning.
\newblock In \emph{ICLR}.

\bibitem[{Li et~al.(2020{\natexlab{b}})Li, Huang, Yang, Wang, and
  Zhang}]{li2019convergence}
Li, X.; Huang, K.; Yang, W.; Wang, S.; and Zhang, Z. 2020{\natexlab{b}}.
\newblock On the convergence of fedavg on non-iid data.
\newblock In \emph{ICLR}.

\bibitem[{Li et~al.(2020{\natexlab{c}})Li, Kovalev, Qian, and
  Richt{\'a}rik}]{li2020acceleration}
Li, Z.; Kovalev, D.; Qian, X.; and Richt{\'a}rik, P. 2020{\natexlab{c}}.
\newblock Acceleration for Compressed Gradient Descent in Distributed and
  Federated Optimization.
\newblock In \emph{ICML}.

\bibitem[{Liu et~al.(2020)Liu, Wu, Ge, Fan, and Zou}]{liu2020federated}
Liu, F.; Wu, X.; Ge, S.; Fan, W.; and Zou, Y. 2020.
\newblock Federated Learning for Vision-and-Language Grounding Problems.
\newblock In \emph{AAAI}.

\bibitem[{Malinovsky et~al.(2020)Malinovsky, Kovalev, Gasanov, Condat, and
  Richtarik}]{malinovsky2020local}
Malinovsky, G.; Kovalev, D.; Gasanov, E.; Condat, L.; and Richtarik, P. 2020.
\newblock From Local SGD to Local Fixed Point Methods for Federated Learning.
\newblock In \emph{ICML}.

\bibitem[{McMahan et~al.(2017)McMahan, Moore, Ramage, Hampson
  et~al.}]{mcmahan2016communication}
McMahan, H.~B.; Moore, E.; Ramage, D.; Hampson, S.; et~al. 2017.
\newblock Communication-efficient learning of deep networks from decentralized
  data.
\newblock In \emph{AISTATS}.

\bibitem[{Melis et~al.(2019)Melis, Song, De~Cristofaro, and
  Shmatikov}]{melis2019exploiting}
Melis, L.; Song, C.; De~Cristofaro, E.; and Shmatikov, V. 2019.
\newblock Exploiting unintended feature leakage in collaborative learning.
\newblock In \emph{IEEE S\&P}.

\bibitem[{Mhamdi, Guerraoui, and Rouault(2018)}]{Mhamdi18}
Mhamdi, E. M.~E.; Guerraoui, R.; and Rouault, S. 2018.
\newblock The Hidden Vulnerability of Distributed Learning in Byzantium.
\newblock In \emph{ICML}.

\bibitem[{Mohri, Sivek, and Suresh(2019)}]{mohri2019agnostic}
Mohri, M.; Sivek, G.; and Suresh, A.~T. 2019.
\newblock Agnostic Federated Learning.
\newblock In \emph{ICML}.

\bibitem[{Peng et~al.(2020)Peng, Huang, Zhu, and Saenko}]{peng2019federated}
Peng, X.; Huang, Z.; Zhu, Y.; and Saenko, K. 2020.
\newblock Federated Adversarial Domain Adaptation.
\newblock In \emph{ICLR}.

\bibitem[{Rothchild et~al.(2020)Rothchild, Panda, Ullah, Ivkin, Stoica,
  Braverman, Gonzalez, and Arora}]{rothchildfetchsgd}
Rothchild, D.; Panda, A.; Ullah, E.; Ivkin, N.; Stoica, I.; Braverman, V.;
  Gonzalez, J.; and Arora, R. 2020.
\newblock FetchSGD: Communication-Efficient Federated Learning with Sketching.
\newblock In \emph{ICML}.

\bibitem[{Sahu et~al.(2018)Sahu, Li, Sanjabi, Zaheer, Talwalkar, and
  Smith}]{sahu2018convergence}
Sahu, A.~K.; Li, T.; Sanjabi, M.; Zaheer, M.; Talwalkar, A.; and Smith, V.
  2018.
\newblock On the convergence of federated optimization in heterogeneous
  networks.
\newblock \emph{arXiv preprint arXiv:1812.06127} .

\bibitem[{Smith et~al.(2017)Smith, Chiang, Sanjabi, and
  Talwalkar}]{smith2017federated}
Smith, V.; Chiang, C.-K.; Sanjabi, M.; and Talwalkar, A.~S. 2017.
\newblock Federated multi-task learning.
\newblock In \emph{NeurIPS}.

\bibitem[{Vogels, Karimireddy, and Jaggi(2019)}]{vogels2019powersgd}
Vogels, T.; Karimireddy, S.~P.; and Jaggi, M. 2019.
\newblock PowerSGD: Practical low-rank gradient compression for distributed
  optimization.
\newblock In \emph{NeurIPS}.

\bibitem[{Wang et~al.(2020)Wang, Yurochkin, Sun, Papailiopoulos, and
  Khazaeni}]{wang2020federated}
Wang, H.; Yurochkin, M.; Sun, Y.; Papailiopoulos, D.; and Khazaeni, Y. 2020.
\newblock Federated Learning with Matched Averaging.
\newblock In \emph{ICLR}.

\bibitem[{Wang, Tong, and Shi(2020)}]{wang2020federatedlatent}
Wang, Y.; Tong, Y.; and Shi, D. 2020.
\newblock Federated Latent Dirichlet Allocation: A Local Differential Privacy
  Based Framework.
\newblock In \emph{AAAI}.

\bibitem[{Wen et~al.(2017)Wen, Xu, Yan, Wu, Wang, Chen, and
  Li}]{wen2017terngrad}
Wen, W.; Xu, C.; Yan, F.; Wu, C.; Wang, Y.; Chen, Y.; and Li, H. 2017.
\newblock Terngrad: Ternary gradients to reduce communication in distributed
  deep learning.
\newblock In \emph{NeurIPS}.

\bibitem[{Xie et~al.(2020)Xie, Huang, Chen, and Li}]{xie2019dba}
Xie, C.; Huang, K.; Chen, P.-Y.; and Li, B. 2020.
\newblock DBA: Distributed Backdoor Attacks against Federated Learning.
\newblock In \emph{ICLR}.

\bibitem[{Xie, Koyejo, and Gupta(2019)}]{Xie19}
Xie, C.; Koyejo, S.; and Gupta, I. 2019.
\newblock Fall of empires: Breaking byzantine-tolerant SGD by inner product
  manipulation.
\newblock In \emph{UAI}.

\bibitem[{Yin et~al.(2019)Yin, Chen, Kannan, and Bartlett}]{yin2019defending}
Yin, D.; Chen, Y.; Kannan, R.; and Bartlett, P. 2019.
\newblock Defending Against Saddle Point Attack in Byzantine-Robust Distributed
  Learning.
\newblock In \emph{ICML}.

\bibitem[{Yin et~al.(2018)Yin, Chen, Ramchandran, and Bartlett}]{Yin18}
Yin, D.; Chen, Y.; Ramchandran, K.; and Bartlett, P. 2018.
\newblock Byzantine-Robust Distributed Learning: Towards Optimal Statistical
  Rates.
\newblock In \emph{ICML}.

\bibitem[{Yurochkin et~al.(2019)Yurochkin, Agarwal, Ghosh, Greenewald, Hoang,
  and Khazaeni}]{yurochkin2019bayesian}
Yurochkin, M.; Agarwal, M.; Ghosh, S.; Greenewald, K.; Hoang, N.; and Khazaeni,
  Y. 2019.
\newblock Bayesian Nonparametric Federated Learning of Neural Networks.
\newblock In \emph{ICML}.

\bibitem[{Zhu, Liu, and Han(2019)}]{zhu2019deep}
Zhu, L.; Liu, Z.; and Han, S. 2019.
\newblock Deep leakage from gradients.
\newblock In \emph{NeurIPS}.

\end{thebibliography}

\onecolumn
\appendix

\newpage 

\begin{table}[!h]
\centering
\begin{tabular}{|c|c|} \hline 
{Layer} & {Size} \\ \hline
{Input} & { $28\times28\times1$}\\ \hline
{Convolution + ReLU} & { $5\times5\times20$}\\ \hline
{Max Pooling} & { $2\times2$}\\ \hline
{Convolution + ReLU} & { $5\times5\times50$}\\ \hline
{Max Pooling} & { $2\times2$}\\ \hline
{Fully Connected + ReLU} & {512}\\ \hline
{Softmax} & {10}\\ \hline
\end{tabular}
\caption{The CNN architecture for MNIST.}
\label{tab:cnn}
\end{table}

\section{Proof of Theorem 1}
\label{proof_of_certified_radius}

\begin{figure*}[!h]
    \center
    {\includegraphics[width=0.3\textwidth]{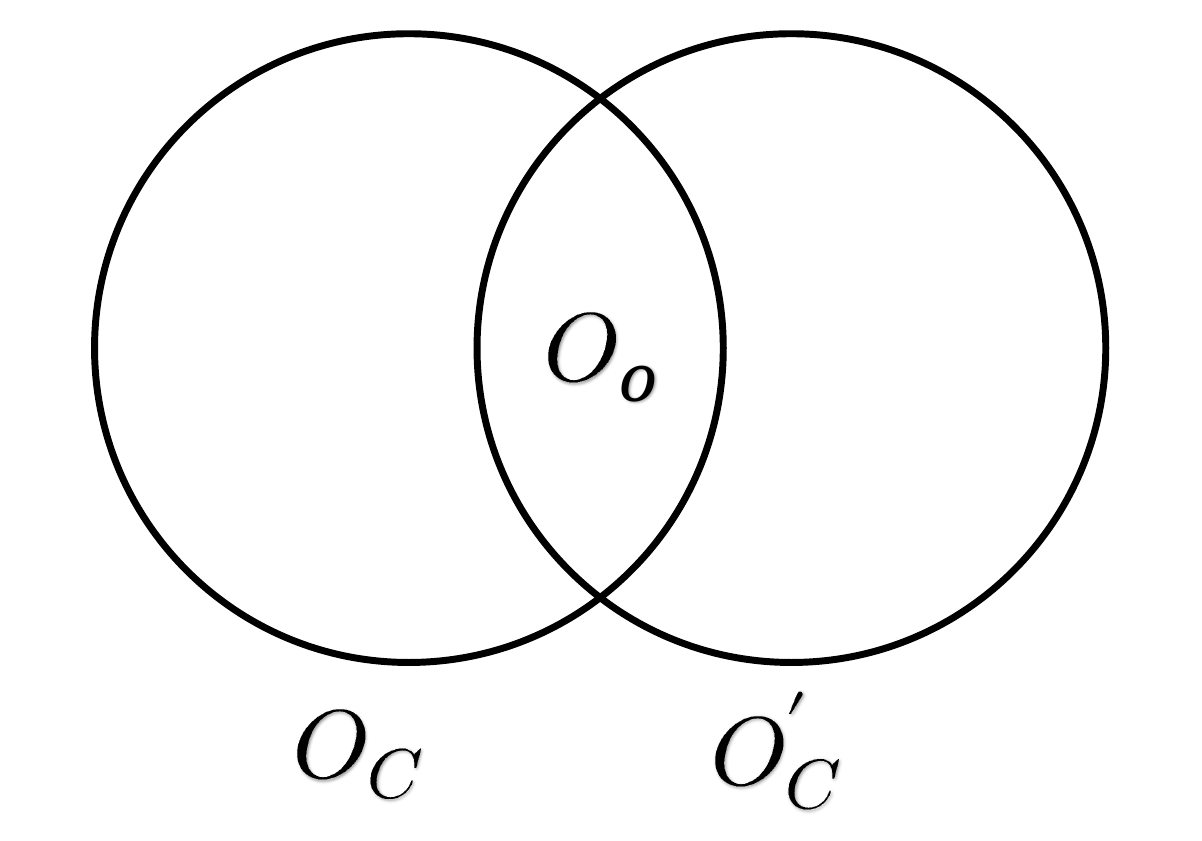}}
    \caption{Illustration of $\mathbf{O}_C,\mathbf{O}_C'$, and $\mathbf{O}_o$.}
    \label{fig:proof1}
\end{figure*}

We first define a subsample of $k$ clients from $\mathbf{C}$ as $\mathcal{S}(\mathbf{C}, k)$. Then, we define the space of all possible subsamples from $\mathbf{C}$ as $\mathbf{O}_C=\{\mathcal{S}(\mathbf{C}, k)\}$ and the space of all possible subsamples from $\mathbf{C}'$ as $\mathbf{O}_C'=\{\mathcal{S}(\mathbf{C}', k)\}$. Let $\mathbf{O}_o=\{\mathcal{S}(\mathbf{C}\cap\mathbf{C}', k)\}=\mathbf{O}_C\cap\mathbf{O}_C'$ denote the space of all possible subsamples from the set of normal clients $\mathbf{C}\cap\mathbf{C}'$, and $\mathbf{O} = \{\mathcal{S}(\mathbf{C}\cup\mathbf{C}', k)\} = \mathbf{O}_C\cup\mathbf{O}_C'$ denote the space of all possible subsamples from either $\mathbf{C}$ or $\mathbf{C}'$. Figure \ref{fig:proof1} illustrates $\mathbf{O}_C, \mathbf{O}_C'$, and $\mathbf{O}_o$. We use a random variable $\mathbf{X}$ to denote a subsample $\mathcal{S}(\mathbf{C}, k)$ and $\mathbf{Y}$ to denote a subsample $\mathcal{S}(\mathbf{C}', k)$ in $\mathbf{O}$. We know that $\mathbf{X}$ and $\mathbf{Y}$ have the following probability distributions:

\begin{align}
	&\text{Pr}(\mathbf{X}=s) = 
		\begin{cases}
 			\frac{1}{{n \choose k}}, &\text{ if } s \in \mathbf{O}_C\\
			0, 					   &\text{ otherwise},
		\end{cases}\\
	&\text{Pr}(\mathbf{Y}=s) = 
		\begin{cases}
 			\frac{1}{{n \choose k}}, &\text{ if } s \in \mathbf{O}_C'\\
			0, 					   &\text{ otherwise}.
		\end{cases}
\end{align}

Recall that given a set of clients $s$, the base federated learning algorithm $\mathcal{A}$ learns a global model. For simplicity, we denote by $\mathcal{A}(s,\mathbf{x})$ the predicted label of a testing example $\mathbf{x}$ given by this global model. 
We have the following equations:

\begin{align}
	p_y &= \text{Pr}(\mathcal{A}(\mathbf{X},\mathbf{x})=y)\\
		&= \text{Pr}(\mathcal{A}(\mathbf{X},\mathbf{x})=y|\mathbf{X}\in\mathbf{O}_o) \cdot \text{Pr}(\mathbf{X}\in\mathbf{O}_o) \nonumber\\\label{eq:subtract1}
		&\;+  \text{Pr}(\mathcal{A}(\mathbf{X},\mathbf{x})=y|\mathbf{X}\in(\mathbf{O}_C-\mathbf{O}_o)) \cdot \text{Pr}(\mathbf{X}\in(\mathbf{O}_C-\mathbf{O}_o)),\\
	p_y' &= \text{Pr}(\mathcal{A}(\mathbf{Y},\mathbf{x})=y)\\\label{eq:subtract2}
		&= \text{Pr}(\mathcal{A}(\mathbf{Y},\mathbf{x})=y|\mathbf{Y}\in\mathbf{O}_o) \cdot \text{Pr}(\mathbf{Y}\in\mathbf{O}_o) \nonumber\\
		&\;+  \text{Pr}(\mathcal{A}(\mathbf{Y},\mathbf{x})=y|\mathbf{Y}\in(\mathbf{O}_C'-\mathbf{O}_o)) \cdot \text{Pr}(\mathbf{Y}\in(\mathbf{O}_C'-\mathbf{O}_o)).
\end{align}

Note that we have:

\begin{align}
	&\text{Pr}(\mathcal{A}(\mathbf{X},\mathbf{x})=y|\mathbf{X}\in\mathbf{O}_o) = \text{Pr}(\mathcal{A}(\mathbf{Y},\mathbf{x})=y|\mathbf{Y}\in\mathbf{O}_o),\\
	&\text{Pr}(\mathbf{X}\in\mathbf{O}_o) = \text{Pr}(\mathbf{Y}\in\mathbf{O}_o) = \frac{{n-m \choose k}}{{n \choose k}}, 
\end{align}

where $m$ is the number of malicious clients. Therefore, we know:

\begin{align}
	\text{Pr}(\mathcal{A}(\mathbf{X},\mathbf{x})=y|\mathbf{X}\in\mathbf{O}_o) \cdot \text{Pr}(\mathbf{X}\in\mathbf{O}_o) = \text{Pr}(\mathcal{A}(\mathbf{Y},\mathbf{x})=y|\mathbf{Y}\in\mathbf{O}_o) \cdot \text{Pr}(\mathbf{Y}\in\mathbf{O}_o).
\end{align}

By subtracting (\ref{eq:subtract1}) from (\ref{eq:subtract2}), we obtain:

\begin{align}
	p_y' - p_y &=  \text{Pr}(\mathcal{A}(\mathbf{Y},\mathbf{x})=y|\mathbf{Y}\in(\mathbf{O}_C'-\mathbf{O}_o)) \cdot \text{Pr}(\mathbf{Y}\in(\mathbf{O}_C'-\mathbf{O}_o))\nonumber\\
			&\;- \text{Pr}(\mathcal{A}(\mathbf{X},\mathbf{x})=y|\mathbf{X}\in(\mathbf{O}_C-\mathbf{O}_o)) \cdot \text{Pr}(\mathbf{X}\in(\mathbf{O}_C-\mathbf{O}_o)).
\end{align}

Similarly, we have the following equation for any $i\neq y$:

\begin{align}
	p_i' - p_i &=  \text{Pr}(\mathcal{A}(\mathbf{Y},\mathbf{x})=i|\mathbf{Y}\in(\mathbf{O}_C'-\mathbf{O}_o)) \cdot \text{Pr}(\mathbf{Y}\in(\mathbf{O}_C'-\mathbf{O}_o))\nonumber\\
			&\;- \text{Pr}(\mathcal{A}(\mathbf{X},\mathbf{x})=i|\mathbf{X}\in(\mathbf{O}_C-\mathbf{O}_o)) \cdot \text{Pr}(\mathbf{X}\in(\mathbf{O}_C-\mathbf{O}_o)).
\end{align}

Therefore, we can show:

\begin{align}
	p_y'- p_i' &= p_y - p_i + (p_y' - p_y) - (p_i' - p_i) \\
			&= p_y - p_i \nonumber\\
				&+  \left[\text{Pr}(\mathcal{A}(\mathbf{Y},\mathbf{x})=y|\mathbf{Y}\in(\mathbf{O}_C'-\mathbf{O}_o)) - \text{Pr}(\mathcal{A}(\mathbf{Y},\mathbf{x})=i|\mathbf{Y}\in(\mathbf{O}_C'-\mathbf{O}_o))\right] \nonumber \cdot \text{Pr}(\mathbf{Y}\in(\mathbf{O}_C'-\mathbf{O}_o))\nonumber\\
				&- \left[\text{Pr}(\mathcal{A}(\mathbf{X},\mathbf{x})=y|\mathbf{X}\in(\mathbf{O}_C-\mathbf{O}_o)) - \text{Pr}(\mathcal{A}(\mathbf{X},\mathbf{x})=i|\mathbf{X}\in(\mathbf{O}_C-\mathbf{O}_o))\right] \cdot \text{Pr}(\mathbf{X}\in(\mathbf{O}_C-\mathbf{O}_o)).\label{eq:difference}
\end{align}

Note that we have:

\begin{align}
	&\text{Pr}(\mathcal{A}(\mathbf{Y},\mathbf{x})=y|\mathbf{Y}\in(\mathbf{O}_C'-\mathbf{O}_o)) - \text{Pr}(\mathcal{A}(\mathbf{Y},\mathbf{x})=i|\mathbf{Y}\in(\mathbf{O}_C'-\mathbf{O}_o)) \ge -1, \\
	&\text{Pr}(\mathcal{A}(\mathbf{X},\mathbf{x})=y|\mathbf{X}\in(\mathbf{O}_C-\mathbf{O}_o)) - \text{Pr}(\mathcal{A}(\mathbf{X},\mathbf{x})=i|\mathbf{X}\in(\mathbf{O}_C-\mathbf{O}_o)) \le 1,\\
	&\text{Pr}(\mathbf{Y}\in(\mathbf{O}_C'-\mathbf{O}_o)) = \text{Pr}(\mathbf{X}\in(\mathbf{O}_C-\mathbf{O}_o)) = 1 - \frac{{n-m \choose k}}{{n \choose k}}. 
\end{align}

 Therefore, based on (\ref{eq:difference}) and that $p_y$ and $p_i$ are integer multiplications of $\frac{1}{{n \choose k}}$, we have the following: 

\begin{align}
	p_y'- p_i' &\ge p_y - p_i + (-1)\cdot \left[1 - \frac{{n-m \choose k}}{{n \choose k}}\right] - \left[1 - \frac{{n-m \choose k}}{{n \choose k}}\right]\\
			&= p_y - p_i - \left[2 - 2\cdot\frac{{n-m \choose k}}{{n \choose k}}\right]\\
			&= \frac{\left\lceil p_y \cdot {n \choose k}\right\rceil}{{n \choose k}} - \frac{\left\lfloor p_i \cdot {n \choose k}\right\rfloor}{{n\choose k}} - 2\left[1 - \frac{{n-m \choose k}}{{n \choose k}}\right] \\
			&\ge \frac{\left\lceil\underline{p_y} \cdot {n \choose k}\right\rceil}{{n \choose k}} - \frac{\left\lfloor\overline{p}_z \cdot {n \choose k}\right\rfloor}{{n\choose k}} - 2\left[1 - \frac{{n-m^* \choose k}}{{n \choose k}}\right]\\
			&> 0,
\end{align}

which indicates $h(\mathbf{C'}, \mathbf{x}) = y$. 

\section{Proof of Theorem 2}
\label{proof_of_tightness}

We prove Theorem~\ref{tightness_theorem} by constructing a base federated learning algorithm $\mathcal{A}^*$ 
such that the conditions in (\ref{eq:prob_condition}) are satisfied but $h(\mathbf{C'}, \mathbf{x}) \neq y$ or there exist ties. 

\begin{figure*}[!t]
    \center
    \subfigure[Case 1]{\includegraphics[width=0.45\textwidth]{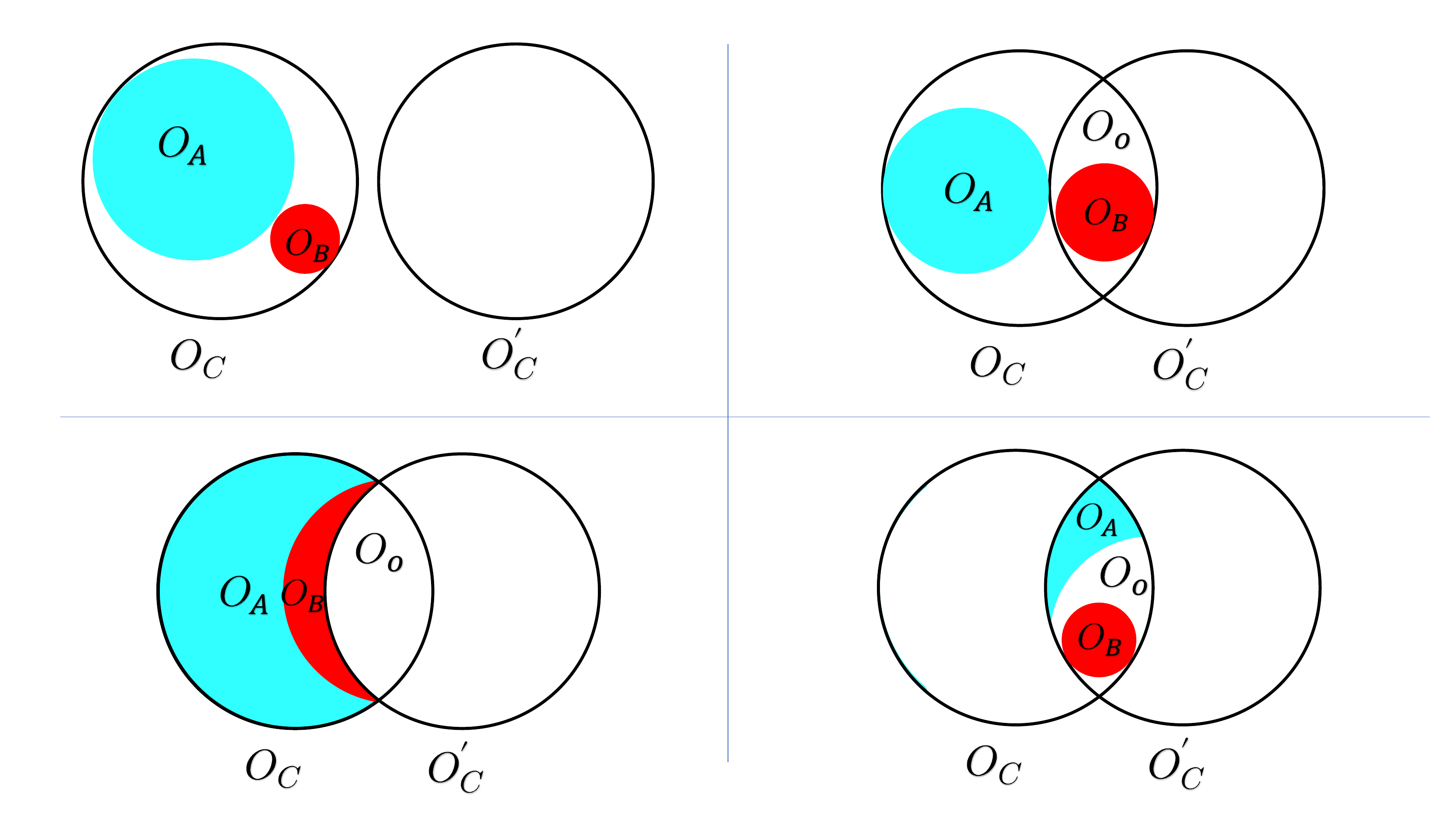}\label{fig:case1}}
    \subfigure[Case 2]{\includegraphics[width=0.45\textwidth]{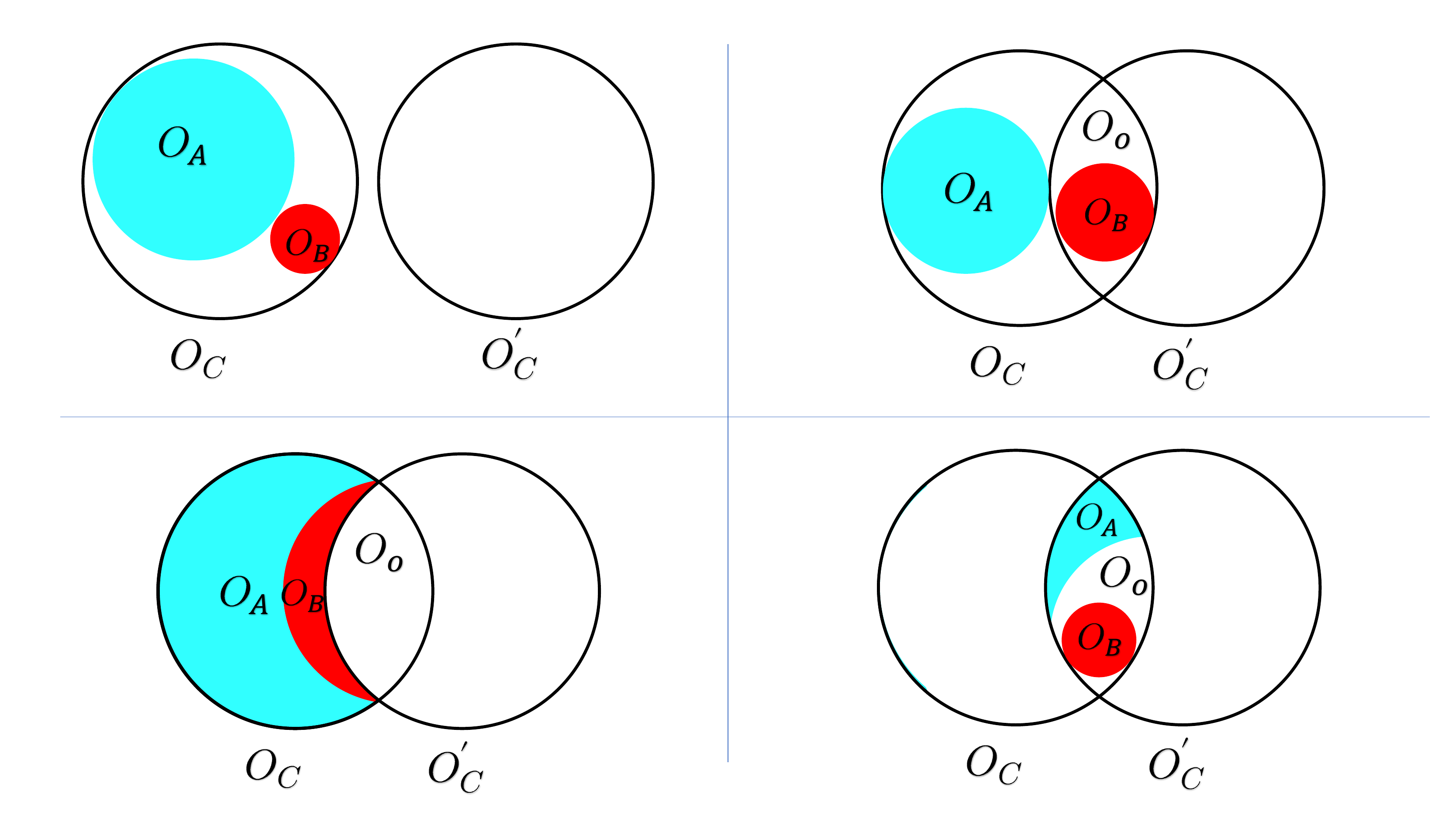}\label{fig:case2}}\\
    \subfigure[Case 3]{\includegraphics[width=0.45\textwidth]{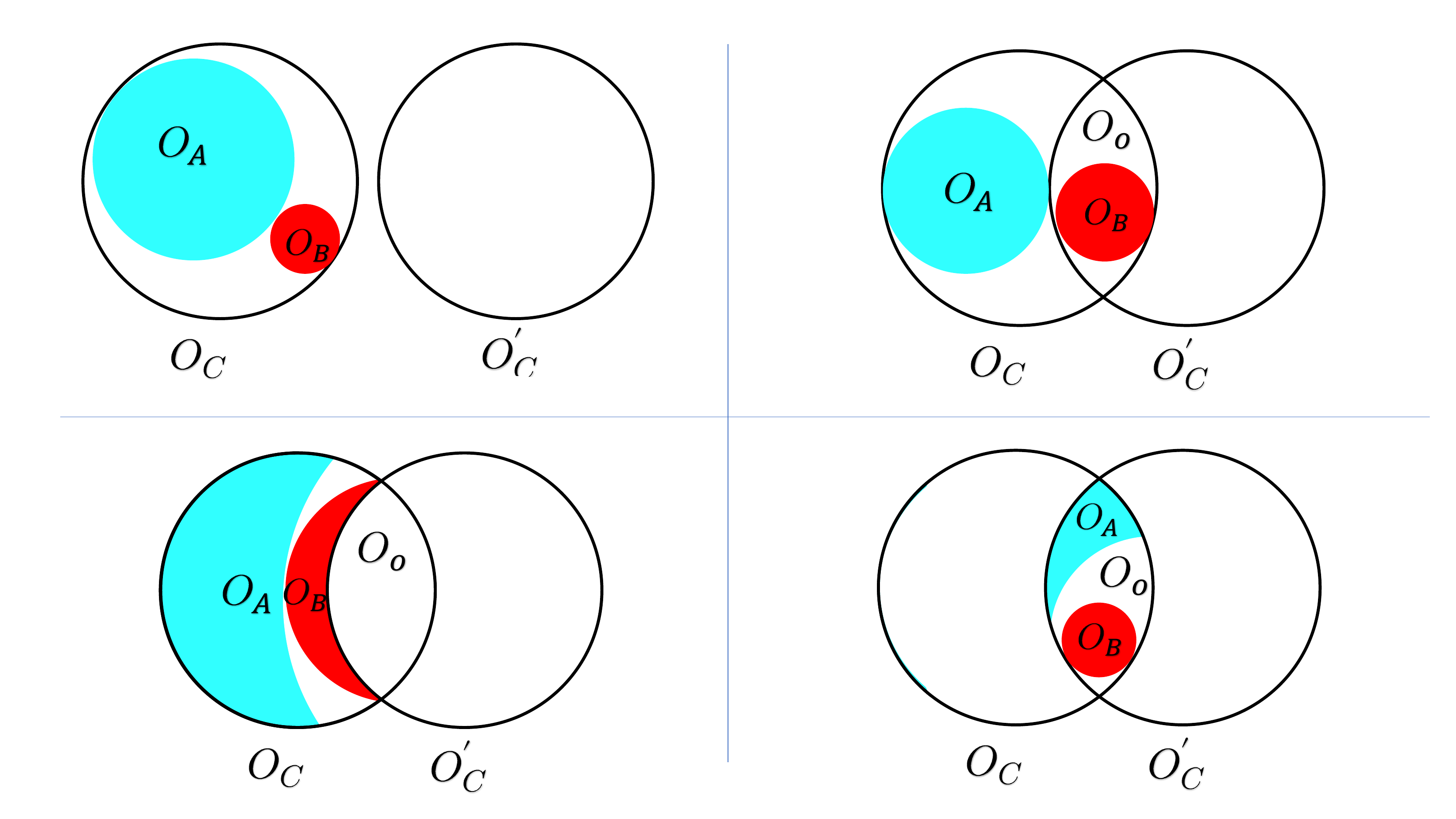}\label{fig:case3}}
    \subfigure[Case 4]{\includegraphics[width=0.45\textwidth]{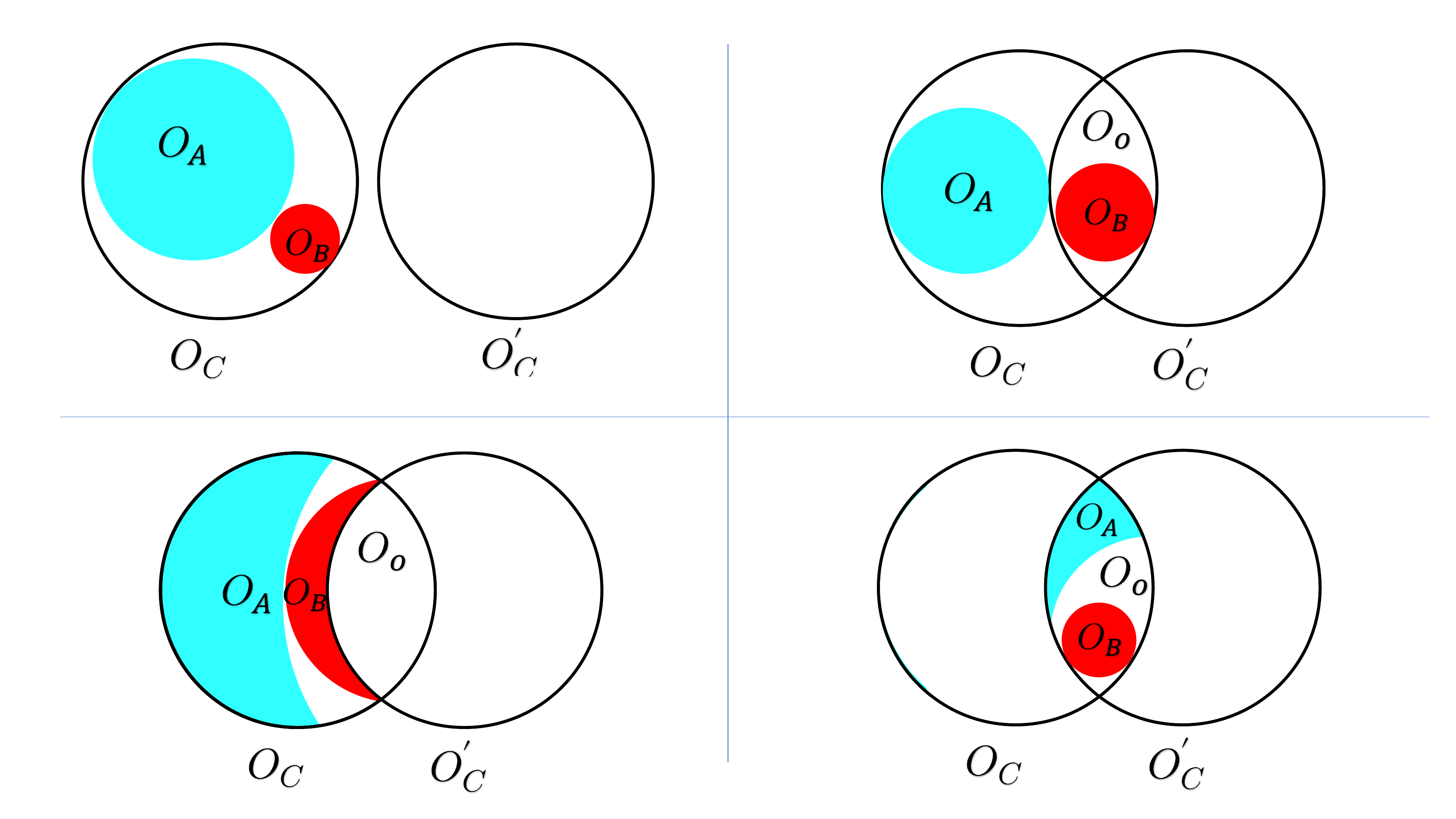}\label{fig:case4}}
    \caption{Illustration of $\mathbf{O}_C, \mathbf{O}_C', \mathbf{O}_o, \mathbf{O}_A,$ and $\mathbf{O}_B$ in the four cases.} 
    \label{fig:four_cases}
\end{figure*}

We follow the definitions of $\mathbf{O}, \mathbf{O}_C, \mathbf{O}_C', \mathbf{O}_o, \mathbf{X}$, and $\mathbf{Y}$ in the previous section.
Next, we consider four cases (Figure \ref{fig:four_cases} illustrates them).

\begin{enumerate}[label={\bfseries Case \arabic*:}, wide=0pt]

\item $m\ge n-k$.

In this case, we know $\mathbf{O}_o=\emptyset$.
Let $\mathbf{O}_A \subseteq \mathbf{O}_C$ and $\mathbf{O}_B \subseteq \mathbf{O}_C$ such that $|\mathbf{O}_A|=\left\lceil\underline{p_y} \cdot {n \choose k}\right\rceil$, $|\mathbf{O}_B|=\left\lfloor\overline{p}_z \cdot {n \choose k}\right\rfloor$, and $\mathbf{O}_A\cap\mathbf{O}_B=\emptyset$. Since $\underline{p_y} + \overline{p}_z \leq 1$, we have:

\begin{align}
	|\mathbf{O}_A| + |\mathbf{O}_B| &= \left\lceil\underline{p_y} \cdot {n \choose k}\right\rceil + \left\lfloor\overline{p}_z \cdot {n \choose k}\right\rfloor \\
									&\le \left\lceil\underline{p_y} \cdot {n \choose k}\right\rceil + \left\lfloor(1-\underline{p_y})\cdot {n \choose k}\right\rfloor \\
									&= \left\lceil\underline{p_y} \cdot {n \choose k}\right\rceil +  {n \choose k} - \left\lceil\underline{p_y} \cdot {n \choose k}\right\rceil \\
									&= {n \choose k}=|\mathbf{O}_C|.
\end{align}

Therefore, we can always find such a pair of disjoint sets $(\mathbf{O}_A, \mathbf{O}_B)$. Figure \ref{fig:case1} illustrates $\mathbf{O}_A, \mathbf{O}_B,\mathbf{O}_C$, and $\mathbf{O}_C'$.
We can construct $\mathcal{A}^*$ as follows:

\begin{align}
		\mathcal{A}^*(s,\mathbf{x})=
		\begin{cases}
 			y, &\text{ if } s\in \mathbf{O}_A\\
			z, &\text{ if } s\in \mathbf{O}_B\cup\mathbf{O}_C'\\
 			i, i\neq y \text{ and } i\neq z, &\text{ otherwise}.
		\end{cases}
\end{align}

We can show that such $\mathcal{A}^*$ satisfies the following probability properties:

\begin{align}
	p_y &= \text{Pr}(\mathcal{A}^*(\mathbf{X},\mathbf{x})=y) = \frac{|\mathbf{O}_A|}{|\mathbf{O}_C|} = \frac{\left\lceil\underline{p_y} \cdot {n \choose k}\right\rceil}{{n \choose k}}\ge \underline{p_y},\\
	p_z &= \text{Pr}(\mathcal{A}^*(\mathbf{X},\mathbf{x})=z) = \frac{|\mathbf{O}_B|}{|\mathbf{O}_C|} = \frac{\left\lfloor\overline{p}_z \cdot {n \choose k}\right\rfloor}{{n \choose k}} \le\overline{p}_z.
\end{align}

Therefore, $\mathcal{A}^*$ satisfies the probability conditions in (\ref{eq:prob_condition}). However, we have:

\begin{align}
	p_z' = \text{Pr}(\mathcal{A}^*(\mathbf{Y},\mathbf{x})=z) = 1,
\end{align}

which indicates $h(\mathbf{C'}, \mathbf{x}) = z \neq y$.

 \item $m^* < m < n-k$, $0 \le \underline{p_y} \le 1-\frac{{n-m \choose k}}{{n \choose k}}$, and $0\le \overline{p}_z \le \frac{{n-m \choose k}}{{n \choose k}}$.

Let $\mathbf{O}_A\subseteq \mathbf{O}_C - \mathbf{O}_o$ such that $|\mathbf{O}_A|=\lceil\underline{p_y} \cdot {n \choose k}\rceil$. 
Let $\mathbf{O}_B\subseteq \mathbf{O}_o$ such that $|\mathbf{O}_B|=\lfloor\overline{p}_z \cdot {n \choose k}\rfloor$. 
Figure \ref{fig:case2} illustrates $\mathbf{O}_A, \mathbf{O}_B,\mathbf{O}_C, \mathbf{O}_C'$, and $\mathbf{O}_o$. We can construct a federated 
learning algorithm $\mathcal{A}^*$ as follows:

	\begin{align}
		\mathcal{A}^*(s,\mathbf{x})=
		\begin{cases}
 			y, &\text{ if } s\in \mathbf{O}_A\\
			z, &\text{ if } s\in \mathbf{O}_B\cup(\mathbf{O}_C'-\mathbf{O}_o)\\
 			i, i\neq y \text{ and } i\neq z, &\text{ otherwise}.
		\end{cases}
	\end{align}

	We can show that such $\mathcal{A}^*$ satisfies the following probability conditions:

	\begin{align}
		p_y &= \text{Pr}(\mathcal{A}^*(\mathbf{X},\mathbf{x})=y)= \frac{|\mathbf{O}_A|}{|\mathbf{O}_C|} = \frac{\left\lceil\underline{p_y} \cdot {n \choose k}\right\rceil}{{n \choose k}}\ge \underline{p_y},\\
		p_z &= \text{Pr}(\mathcal{A}^*(\mathbf{X},\mathbf{x})=z) = \frac{|\mathbf{O}_B|}{|\mathbf{O}_C|} = \frac{\left\lfloor\overline{p}_z \cdot {n \choose k}\right\rfloor}{{n \choose k}} \le\overline{p}_z,
	\end{align}

	which indicates $\mathcal{A}^*$ satisfies (\ref{eq:prob_condition}). However, we have:

	\begin{align}
		p_y' - p_z' &= \text{Pr}(\mathcal{A}^*(\mathbf{Y},\mathbf{x})=y) - \text{Pr}(\mathcal{A}^*(\mathbf{Y},\mathbf{x})=z)\\
					&= 0 - \frac{|\mathbf{O}_B|+|\mathbf{O}_C'-\mathbf{O}_o|}{|\mathbf{O}_C'|}\\
					&= -\frac{\left\lfloor\overline{p}_z \cdot {n \choose k}\right\rfloor}{{n\choose k}} - 1 + \frac{{n-m \choose k}}{{n \choose k}}\\
					&< 0, 
	\end{align}
	which implies $h(\mathbf{C'}, \mathbf{x}) \neq y$.

  \item $m^* < m < n-k$, $0\le \underline{p_y} \le 1-\frac{{n-m \choose k}}{{n \choose k}}$, and $\frac{{n-m \choose k}}{{n \choose k}}\le\overline{p}_z \le 1-\underline{p_y}$.

Let $\mathbf{O}_A\subseteq \mathbf{O}_C-\mathbf{O}_o$  and $\mathbf{O}_B\subseteq \mathbf{O}_C-\mathbf{O}_o$ such that $|\mathbf{O}_A|=\lceil\underline{p_y} \cdot {n \choose k}\rceil$, $|\mathbf{O}_B|=\lfloor\overline{p}_z \cdot {n \choose k}\rfloor - {n-m \choose k}$, and $\mathbf{O}_A \cap \mathbf{O}_B=\emptyset$. Note that $|\mathbf{O}_C-\mathbf{O}_o| = {n \choose k}-{n-m \choose k}$, and we have:

\begin{align}
|\mathbf{O}_A|+|\mathbf{O}_B| &= \left\lceil\underline{p_y} \cdot {n \choose k}\right\rceil + \left\lfloor\overline{p}_z \cdot {n \choose k}\right\rfloor - {n-m \choose k} \\
							&\le \left\lceil\underline{p_y} \cdot {n \choose k}\right\rceil + \left\lfloor(1-\underline{p_y}) \cdot {n \choose k}\right\rfloor - {n-m \choose k} \\
							&= \left\lceil\underline{p_y} \cdot {n \choose k}\right\rceil + \left[{n \choose k} - \left\lceil\underline{p_y} \cdot {n \choose k}\right\rceil\right] -{n-m \choose k} \\
							&= {n \choose k} - {n-m \choose k}.
\end{align}

Therefore, we can always find a pair of such disjoint sets $(\mathbf{O}_A, \mathbf{O}_B)$. Figure \ref{fig:case3} illustrates $\mathbf{O}_A, \mathbf{O}_B,\mathbf{O}_C, \mathbf{O}_C'$, and $\mathbf{O}_o$. We can  construct an algorithm $\mathcal{A}^*$ as follows:

	\begin{align}
		\mathcal{A}^*(s,\mathbf{x})=
		\begin{cases}
 			y, &\text{ if } s\in \mathbf{O}_A\\
			z, &\text{ if } s\in \mathbf{O}_B\cup\mathbf{O}_C'\\
 			i, i\neq y \text{ and } i\neq z, &\text{ otherwise}.
		\end{cases}
	\end{align}

	We can show that such $\mathcal{A}^*$ satisfies the following probability conditions:
	\begin{align}
		p_y &= \text{Pr}(\mathcal{A}^*(\mathbf{X},\mathbf{x})=y) = \frac{|\mathbf{O}_A|}{|\mathbf{O}_C|} = \frac{\left\lceil\underline{p_y} \cdot {n \choose k}\right\rceil}{{n \choose k}}\ge \underline{p_y},\\
		p_z &= \text{Pr}(\mathcal{A}^*(\mathbf{X},\mathbf{x})=z) = \frac{|\mathbf{O}_B|+|\mathbf{O}_o|}{|\mathbf{O}_C|} = \frac{\left\lfloor\overline{p}_z \cdot {n \choose k}\right\rfloor}{{n \choose k}} \le\overline{p}_z,
	\end{align}
	which are consistent with the probability conditions in (\ref{eq:prob_condition}). However, we can show the following:
	\begin{align}
		 p_z' =  \text{Pr}(\mathcal{A}^*(\mathbf{Y},\mathbf{x})=z)=1,
	\end{align}
	which gives $h(\mathbf{C'}, \mathbf{x}) = z \neq y$.

  \item $m^* < m < n-k$, $1-\frac{{n-m \choose k}}{{n \choose k}}< \underline{p_y} \le 1$, and $0\leq \overline{p}_z \le 1 - \underline{p_y}<\frac{{n-m \choose k}}{{n \choose k}}$.

Let $\mathbf{O}_A\subseteq \mathbf{O}_o$ and $\mathbf{O}_B\subseteq \mathbf{C}_o$ such that $|\mathbf{O}_A|=\left\lceil\underline{p_y} \cdot {n \choose k}\right\rceil + {n-m \choose k} - {n \choose k}$, $|\mathbf{O}_B|=\left\lfloor\overline{p}_z \cdot {n \choose k}\right\rfloor$, and $\mathbf{O}_A \cap \mathbf{O}_B=\emptyset$. Note that $|\mathbf{O}_o| = {n-m \choose k}$, and we have:
\begin{align}
|\mathbf{O}_A|+|\mathbf{O}_B| &= \left\lceil\underline{p_y} \cdot {n \choose k}\right\rceil + {n-m \choose k} - {n \choose k} + \left\lfloor\overline{p}_z \cdot {n \choose k}\right\rfloor \\
							&\le \left\lceil\underline{p_y} \cdot {n \choose k}\right\rceil + {n-m \choose k} - {n \choose k} + \left\lfloor(1-\underline{p_y}) \cdot {n \choose k}\right\rfloor \\
							&= \left\lceil\underline{p_y} \cdot {n \choose k}\right\rceil + {n-m \choose k} - {n \choose k} + \left[{n \choose k} - \left\lceil\underline{p_y} \cdot {n \choose k}\right\rceil\right]  \\
							&= {n-m \choose k}.
\end{align}
Therefore, we can always find such a pair of disjoint sets $(\mathbf{O}_A$, $\mathbf{O}_B)$. Figure \ref{fig:case4} illustrates $\mathbf{O}_A, \mathbf{O}_B,\mathbf{O}_C, \mathbf{O}_C'$, and $\mathbf{O}_o$. Next, we can construct an algorithm $\mathcal{A}^*$ as follows:
	\begin{align}
		\mathcal{A}^*(s,\mathbf{x})=
		\begin{cases}
 			y, &\text{ if } s\in \mathbf{O}_A\cup(\mathbf{O}_C - \mathbf{O}_o)\\
			z, &\text{ if } s\in \mathbf{O}_B\cup(\mathbf{O}_C' - \mathbf{O}_o)\\
 			i, i\neq y \text{ and } i\neq z, &\text{ otherwise}.
		\end{cases}
	\end{align}
	We can show that $\mathcal{A}^*$ has the following properties:
	\begin{align}
		p_y &= \text{Pr}(\mathcal{A}^*(\mathbf{X},\mathbf{x})=y) = \frac{|\mathbf{O}_A|+|\mathbf{O}_C-\mathbf{O}_o|}{|\mathbf{O}_C|} = \frac{\left\lceil\underline{p_y} \cdot {n \choose k}\right\rceil}{{n \choose k}}\ge \underline{p_y},\\
		p_z &= \text{Pr}(\mathcal{A}^*(\mathbf{X},\mathbf{x})=z) = \frac{|\mathbf{O}_B|}{|\mathbf{O}_C|} = \frac{\left\lfloor\overline{p}_z \cdot {n \choose k}\right\rfloor}{{n \choose k}} \le\overline{p}_z,
	\end{align}
	which implies $\mathcal{A}^*$ satisfies the probability conditions in (\ref{eq:prob_condition}). However, we  also have:
	\begin{align}
		p_y' - p_z' &= \text{Pr}(\mathcal{A}^*(\mathbf{Y},\mathbf{x})=y) -  \text{Pr}(\mathcal{A}^*(\mathbf{Y},\mathbf{x})=z)\\
					& = \frac{|\mathbf{O}_A|}{|\mathbf{O}_C'|} - \frac{|\mathbf{O}_B|+|\mathbf{O}_C'-\mathbf{O}_o|}{|\mathbf{O}_C'|}\\
					&= \frac{\left\lceil\underline{p_y} \cdot {n \choose k}\right\rceil + {n-m \choose k} - {n \choose k}}{{n \choose k}} - \frac{\left\lfloor\overline{p}_z \cdot {n \choose k}\right\rfloor - {n-m \choose k} + {n \choose k}}{{n \choose k}}\\
					&= \frac{\left\lceil\underline{p_y} \cdot {n \choose k}\right\rceil}{{n \choose k}} - \frac{\left\lfloor\overline{p}_z \cdot {n \choose k}\right\rfloor}{{n\choose k}} - \left[2 - 2\cdot \frac{{n-m \choose k}}{{n \choose k}}\right].
	\end{align}
	 Since $m>m^*$, we have:
	\begin{align}
	\frac{\left\lceil\underline{p_y} \cdot {n \choose k}\right\rceil}{{n \choose k}} - \frac{\left\lfloor\overline{p}_z \cdot {n \choose k}\right\rfloor}{{n\choose k}}  \le \left[2 - 2\cdot \frac{{n-m \choose k}}{{n \choose k}}\right].
	\end{align}
	Therefore, we have $p_y' - p_z' \le 0$,
	which indicates $h(\mathbf{C'}, \mathbf{x}) \neq y$ or there exist ties.

\end{enumerate}
To summarize, we have proven that in any possible cases, Theorem \ref{tightness_theorem} holds, indicating that our derived certified security level is tight.

\section{Proof of Theorem 3}
\label{proof_of_probability}

Based on the Clopper-Pearson method, for each testing example $\mathbf{x}_t$, we have:  

\begin{align}
    \text{Pr}(\underline{p_{\hat{y}_t}} \leq \text{Pr}(\mathcal{A}(\mathcal{S}(\mathbf{C},k), \mathbf{x}_t)=\hat{y}_t) \land \overline{p}_{\hat{z}_t}\geq \text{Pr}(\mathcal{A}(\mathcal{S}(\mathbf{C},k), \mathbf{x}_t)=i), \forall i \neq \hat{y}_t) \geq 1 - \frac{\alpha}{d}.
\end{align}

Therefore, for a testing example $\mathbf{x}_t$, if our Algorithm \ref{alg:certify} does not abstain for $\mathbf{x}_t$, the probability that it returns an incorrect certified security level is at most $\frac{\alpha}{d}$. Formally, we have the following:  

\begin{align}
    \text{Pr}((\exists \mathbf{C}', M(\mathbf{C}')\leq \hat{m}_t^*, h(\mathbf{C}',\mathbf{x}_t)\neq\hat{y}_t)|\hat{y}_t \neq \text{ABSTAIN} ) \leq \frac{\alpha}{d}. 
\end{align}

Therefore, we have the following: 
\begin{align}
&\quad\text{Pr}(\cap_{\mathbf{x}_t \in \mathcal{D}} ((\forall \mathbf{C}',M(\mathbf{C}')\leq \hat{m}_t^*, h(\mathbf{C}', \mathbf{x}_t)=\hat{y}_t)|\hat{y}_t\neq \text{ABSTAIN})) \\
\label{theorem_4_apply_oooles_inequality_1}
&= 1 - \text{Pr}(\cup_{\mathbf{x}_t \in \mathcal{D}} ((\exists \mathbf{C}', M(\mathbf{C}')\leq \hat{m}_t^*, h(\mathbf{C}',\mathbf{x}_t)\neq\hat{y}_t )|\hat{y}_t\neq \text{ABSTAIN})) \\
\label{theorem_4_apply_oooles_inequality_2}
& \geq  1 - \sum_{\mathbf{x}_t \in \mathcal{D}}\text{Pr}((\exists \mathbf{C}', M(\mathbf{C}')\leq \hat{m}_t^*, h(\mathbf{C}',\mathbf{x}_t)\neq\hat{y}_t )|\hat{y}_t\neq \text{ABSTAIN}) \\
& \geq  1- d \cdot \frac{\alpha}{d} \\
& = 1 -\alpha .
\end{align}
We have (\ref{theorem_4_apply_oooles_inequality_2}) from (\ref{theorem_4_apply_oooles_inequality_1}) based on the Boole's inequality. 

\end{document}